\documentclass[a4paper,11pt]{article}
\usepackage[utf8]{inputenc}
\usepackage{xcolor}
\usepackage{hyperref}
\definecolor{linkcolor}{HTML}{799B03}
\definecolor{urlcolor}{HTML}{799B03}
\hypersetup{pdfstartview=FitH,linkcolor=linkcolor,urlcolor=urlcolor,colorlinks=true}
\usepackage[english]{babel}
\usepackage{amsmath, amsthm, amsfonts, amssymb,amscd}
\usepackage[all]{xy}
\usepackage{graphicx}
\usepackage{import}
\usepackage{xifthen}
\usepackage{pdfpages}
\usepackage{float}
\usepackage{transparent}
\usepackage{multirow}
\usepackage{cellspace}
\usepackage{colortbl}
\usepackage{bbm}
\usepackage{mathrsfs}
\usepackage{csquotes}
\graphicspath{{pic/}}
\def\[{\begin{equation}}
\def\]{\end{equation}}
\def\X{\mathcal{X}}
\def\Y{\mathcal{Y}}
\def\Z{\mathcal{Z}} 
\DeclareUnicodeCharacter{0301}{`}

\begin{document}
\begin{center}
{\LARGE \bf Stable cosmological solutions in Horndeski theory}

\vspace{20pt}

S. Mironov$^{a,b,c}$\footnote{sa.mironov\_1@physics.msu.ru},
A. Shtennikova$^{a,d}$\footnote{shtennikova@inr.ru}

\renewcommand*{\thefootnote}{\arabic{footnote}}
\vspace{15pt}

$^a$\textit{Institute for Nuclear Research of the Russian Academy of Sciences,\\
60th October Anniversary Prospect, 7a, 117312 Moscow, Russia}\\
\vspace{5pt}
$^b$\textit{Institute for Theoretical and Mathematical Physics,\\ Lomonosov Moscow State University, Moscow
119991, Russia}\\
$^c$\textit{NRC “Kurchatov Institute”, 123182, Moscow, Russia}\\
$^d$\textit{Department of Particle Physics and Cosmology, Physics Faculty,\\
M.V. Lomonosov Moscow State University,\\
Vorobjevy Gory, 119991 Moscow, Russia}
\end{center}

\vspace{5pt}

\begin{abstract}
It is known that the construction of a completely stable solution in Horndeski theory is restricted very strongly by the so-called no-go theorem. Previously, various techniques have been used to avoid the conditions of the theorem. In this paper a new way of constructing stable solutions are shown in the general Horndeski theory. We considered the situation in which the unitary gauge studied earlier turns out to be singular. On this basis we construct a spatially flat, stable bouncing and genesis Universe solutions which are described by General Relativity with non-conventional scalar field.
\end{abstract}

\section{Introduction and summary}

Horndeski theory \cite{Horndeski:1974wa,Nicolis:2008in,Deffayet:2009wt,Fairlie:1991qe}(See \cite{Kobayashi:2019hrl} for a review) is the most general scalar-tensor theory of gravity and one additional scalar field (we denote it by $\pi$) whose equations of motion do not exceed the second order despite the presence of higher derivatives in the Lagrangian. An extension of Horndeski theory is beyond Horndeski~\cite{Zumalacarregui:2013pma,Gleyzes:2014dya}, in which the third derivatives in the equations of motion arise, but still do not lead to Ostrogradsky instability. Futher generalization -- DHOST theories\cite{Langlois:2015cwa,BenAchour:2016fzp,Langlois:2017mxy} (See for a review~\cite{Langlois:2018dxi}).

Since Horndeski theories were rediscovered in 2010, interest in their study has only grown. This is due, among other things, to the possibility of constructing interesting cosmological solutions for scenarios of the early Universe, such as bounce and genesis\cite{Creminelli:2006xe,Kobayashi:2010cm,Creminelli:2010ba,Kobayashi:2011nu,Koehn:2013upa,Easson:2011zy,Cai:2012va,Pirtskhalava:2014esa,Qiu:2015nha,Kobayashi:2015gga,Wan:2015hya,Koehn:2015vvy,Ijjas:2016tpn}. This possibility is related to the fact that these theories allow one to construct linearly stable solutions that violate the null energy condition\cite{Rubakov:2014jja}.

However, there is a significant restriction on the construction of fully stable cosmological solutions caused by the existence of the so-called no-go theorem\cite{Libanov:2016kfc,Kobayashi:2016xpl}. It was formulated in terms of unitary gauge (that is when perturbation of scalar field $\pi$ is put to 0). The statement of the theorem is that in the absence of strong coupling as well as gradient instabilities and ghosts in quadratic action, it is impossible to construct solution with full evolution on the interval $t = (-\infty, + \infty)$ in General Horndeski theory. However, there is a well-known example of a stable solution existing in any Horndeski theory: the Minkowski space. Still in this case the unitary gauge cannot be chosen due to singularity of action coefficients. Therefore, no-go theorem is not applicable here.

Various approaches to circumvent the no-go theorem have been considered. The first approach is to consider more general beyond Horndesky theory. In this case, the conditions for the stability of the solutions become different, which allows one to avoid the theorem\cite{Cai:2016thi,Creminelli:2016zwa,Kolevatov:2017voe,Cai:2017dyi,Cai:2017tku,Mironov:2018oec,Mironov:2019qjt,Mironov:2019mye,Mironov:2019haz,Ilyas:2020qja, Ilyas:2020zcb, Zhu:2021whu, Zhu:2021ggm,Cai:2022ori}. 

The second approach is to consider the naive strong coupling limit on $+\infty$ or $- \infty$. It has been shown in a number of papers\cite{Kobayashi:2016xpl,Ageeva:2020buc,Ageeva:2020gti,Ageeva:2021yik} that, if we consider the next order of action for perturbations, the strong coupling can be avoided for a specific choice of the parameters of theory.

In this paper we consider the third approach in which using of unitary gauge is not possible due to singularities in the quadratic action. This case should not be confused with $\gamma$-crossing\footnote{One also analyzed the behavior of solutions in the area around singular point of the scalar action coefficients\cite{Mironov:2018oec}. And it has been shown that solutions with General Relativity(GR) in the asymptotics on $\pm \infty$ exist in spite of the appearance of singularities in the linearized equations. This is possible because this singularity does not arise in solutions of the equations, and solution also have flat asymptotics at both infinities in time.} where a solution crosses isolated singular point.

First of all, we make a brief review of perturbations about FLRW background in section~\ref{perturbations}, then we consider the method of obtaining an explicitly gauge-invariant action in section~\ref{gauge_inv_action}, then in section~\ref{A4zero} we consider different variants of the behavior of the theory depending on its parameters. Thus it turns out that in the case when the unitary gauge is singular and the background scalar field is non-trivial (i.e. $\dot{\pi} \neq 0$), the scalar modes of perturbations have no dynamics. On the other hand, if the background scalar field is static and it is Minkowski space, then the scalar modes are regular lorenz invariant waves (with sound speed equal to $c_S^2 = 1$ at high momentum). And finally in section~\ref{solution} we show the particular choices for the action for the case of bounce and genesis cosmology.

It is worth noting that the analysis performed in this paper is valid strictly for the cosmological background. When ones tries to consider less symmetrical spaces, the dynamical modes may arise, including the case of a dynamical background scalar field. This may be due to the difference in wave propagation in different directions of space. However, it may turn out that the velocities of these perturbations are negligibly small. This is a question of subsequent analysis.

\section{Perturbations about FLRW background}\label{perturbations}

We are considering the beyond Horndeski theory with the following Lagrangian:
\begin{subequations}
    \label{lagrangian}
    \[S=\int\mathrm{d}^4x\sqrt{-g}\left(\mathcal{L}_2 + \mathcal{L}_3 + \mathcal{L}_4 + \mathcal{L}_5 + \mathcal{L_{BH}}\right),\]
    \vspace{-0.8cm}
    \begin{align}
    &\mathcal{L}_2=F(\pi,X),\\
    &\mathcal{L}_3=K(\pi,X)\Box\pi,\\
    &\mathcal{L}_4=-G_4(\pi,X)R+2G_{4X}(\pi,X)\left[\left(\Box\pi\right)^2-\pi_{;\mu\nu}\pi^{;\mu\nu}\right],\\
    &\mathcal{L}_5=G_5(\pi,X)G^{\mu\nu}\pi_{;\mu\nu}+\frac{1}{3}G_{5X}\left[\left(\Box\pi\right)^3-3\Box\pi\pi_{;\mu\nu}\pi^{;\mu\nu}+2\pi_{;\mu\nu}\pi^{;\mu\rho}\pi_{;\rho}^{\;\;\nu}\right],\\
    &\mathcal{L_{BH}}=F_4(\pi,X)\epsilon^{\mu\nu\rho}_{\quad\;\sigma}\epsilon^{\mu'\nu'\rho'\sigma}\pi_{,\mu}\pi_{,\mu'}\pi_{;\nu\nu'}\pi_{;\rho\rho'}+
    \\\nonumber&\qquad+F_5(\pi,X)\epsilon^{\mu\nu\rho\sigma}\epsilon^{\mu'\nu'\rho'\sigma'}\pi_{,\mu}\pi_{,\mu'}\pi_{;\nu\nu'}\pi_{;\rho\rho'}\pi_{;\sigma\sigma'},
    \end{align}
\end{subequations} 
where $\pi$ is the scalar field, $X=g^{\mu\nu}\pi_{,\mu}\pi_{,\nu}$, $\pi_{,\mu}=\partial_\mu\pi$, $\pi_{;\mu\nu}=\triangledown_\nu\triangledown_\mu\pi$, $\Box\pi = g^{\mu\nu}\triangledown_\nu\triangledown_\mu\pi$, $G_{4X}=\partial G_4/\partial X$, etc. The Horndeski theory corresponds to $F_4(\pi,X)=F_5(\pi,X)=0$.

In this paper we consider spatially flat FLRW background:
\[ ds^{2} = dt^2 - a^2(t) \left(dx^2 + dy^2 + dz^2\right). \]
The decomposition of metric perturbations $h_{\mu \nu}$ into helicity components in the general case has the form
\begin{subequations}
\begin{align}
&h_{00}=2 \Phi \\
&h_{0 i}= - \partial_{i} \beta + Z_i^T, \\
&h_{i j}=-2 \Psi \delta_{i j}-2 \partial_{i}\partial_{j} E - \left(\partial_{i} W_j^T+\partial_{j} W_i^T\right)+h_{i j},
\end{align}    
\end{subequations}
where $\Phi, \beta, \Psi, E$ - scalar fields, $Z_{i}^{T}, W_{i}^{T}$ - transverse vector fields ($\partial_{i}Z_{i}^{T} = \partial_{i} W_{i}^{T} = 0$), $h_{ij}$ - transverse traceless tensor.

The action for tensor perturbations has the form:
\[\label{tensor_action} S_h^{(2)}=\int \mathrm{d} t \mathrm{~d}^3 x a^3\left[\frac{A_5}{2} \left(\dot{h}_{i j}\right) ^2-\frac{A_2}{a^2}\left( \overrightarrow{\nabla} h_{i j}\right)^2\right].\]
Here dot denotes the derivative with respect to the cosmic time $t$, coefficients $A_i$ are the combinations of the Lagrangian functions and their derivatives. The explicit form of these coefficients can be found in the Appendix A.

The vector perturbation sector turns out to be non-dynamical without additional matter, so we do not consider it.

We mainly concentrate on the consideration of the scalar perturbations. We denote the perturbation of scalar field $\delta \pi = \chi$. Then the second-order action for scalar sector has the form:
\begin{align}\label{action}
    S^{(2)} = \int &\mathrm{d}t\,\mathrm{d}^3x\,a^3 \left({A_{1}}\: {\left(\dot{\Psi}\right)}^{2}+{A_{2}}\: \dfrac{(\overrightarrow{\nabla}\Psi)^2}{a^2}+{A_{3}} \: {\Phi}^{2}+{A_{4}}\: \Phi \dfrac{\overrightarrow{\nabla}^2\beta}{a^2}+{A_{5}}\: \dot{\Psi} \dfrac{\overrightarrow{\nabla}^2\beta}{a^2} +{A_{6}}\: \Phi \dot{\Psi} \right.\nonumber \\
    +{A_{7}}& \: \Phi \dfrac{\overrightarrow{\nabla}^2{\Psi}}{a^2}+{A_{8}} \: \Phi \dfrac{\overrightarrow{\nabla}^2{\chi}}{a^2}+{A_{9}}\: \dfrac{\overrightarrow{\nabla}^2{\beta}}{a^2} \dot{\chi} +{A_{10}} \:\chi\ddot{\Psi}+{A_{11}} \:\Phi \dot{\chi}+{A_{12}}\: \chi \dfrac{\overrightarrow{\nabla}^2\beta}{a^2}+{A_{13}} \:\chi \dfrac{\overrightarrow{\nabla}^2\Psi}{a^2} \nonumber\\
    +{A_{14}}&\: {\left(\dot{\chi}\right)}^{2}+{A_{15}}\: \dfrac{(\overrightarrow{\nabla}\chi)^2}{a^2} + A_{16}\:\dot{\chi}\dfrac{\overrightarrow{\nabla}^2\Psi}{a^2} +{A_{17}}\: \Phi \chi +{A_{18}}\: \chi \dot{\Psi}+{A_{19}}\: \Psi \chi + {A_{20}} \:{\chi}^{2}+{A_{21}} \chi \overrightarrow{\nabla}^2 E\nonumber\\ 
       + {A_{22}}&\:  \left.\ddot{\chi} \overrightarrow{\nabla}^2 E +{A_{23}}\: \dot{\chi} \overrightarrow{\nabla}^2 E +{A_{24}}\: \dot{\Phi} \overrightarrow{\nabla}^2 E  +{A_{25}}\: \Phi \overrightarrow{\nabla}^2 E  +{A_{26}}\:\ddot{\Psi} \overrightarrow{\nabla}^2 E +{A_{27}}\: \dot{\Psi}  \overrightarrow{\nabla}^2 E \right).
\end{align}
Note that $A_{19}$ vanishes on the background equations (the equations are given in Appendix B).

\section{Actions for the scalar perturbation sector in terms of gauge-invariant variables}\label{gauge_inv_action}
\subsection{Action for the scalar perturbations in terms of gauge-invariant variables}\label{thr_var_action}

In this section we will show how it is possible to rewrite the action~\eqref{action} in terms of gauge-invariant variables, which will further facilitate our analysis.

Action~\eqref{action} is invariant with respect to small coordinate transformations:
$$x^{\mu} \rightarrow x^{\mu} - \xi^\mu,$$
where $\xi^{\mu} = \left(\xi_0, \xi^i_T + \delta^{i j} \partial_{j}{\xi_{S}}\right)^\mathrm{T}$. For this transformations, the fields are transformed as:
\[\label{gauge} \Phi \to \Phi + \dot{\xi}_0,\quad \beta \to \beta - \xi_0 + a^2 \dot{\xi}_S, \quad \chi \to
\chi + \xi_0\dot{\pi},\quad \Psi \to \Psi + \xi_0 H, \quad E \rightarrow E - \xi_S.\]
The vector $\xi^i_T$, meanwhile, transforms the vector perturbations.

We can introduce new gauge-invariant variables (Bardeen variables):
\begin{subequations}
\label{gauge_inv_var}
    \begin{align}
    \X &= \chi +  \dot{\pi} \left(\frac{\beta}{a^2} + \dot{E}\right)  ,\\
    \mathcal{Y} &= \Psi + H \left(\frac{\beta}{a^2} + \dot{E}\right) ,\\
    \mathcal{Z} &=  \Phi + \frac{\mathrm{d}}{\mathrm{d}t}\left[\frac{\beta}{a^2} + \dot{E}\right].
\end{align}
\end{subequations}

Then, in terms of these variables, the action~\eqref{action} will take the form
\begin{align}\label{three_var_action}
    S^{(2)} = \int &\mathrm{d}t\,\mathrm{d}^3x\,a^3 \left({A_{1}} \: {\left(\dot{\mathcal{Y}}\right)}^{2} +{A_{2}}\: \dfrac{(\overrightarrow{\nabla}\mathcal{Y})^2}{a^2} + {A_{3}}\: {\mathcal{Z}}^{2} +{A_{6}}\: \mathcal{Z} \dot{\mathcal{Y}}  +{A_{7}} \: \mathcal{Z} \dfrac{\overrightarrow{\nabla}^2 \mathcal{Y}}{a^2} +{A_{8}} \: \mathcal{Z} \dfrac{\overrightarrow{\nabla}^2 \mathcal{X}}{a^2}\right. \nonumber \\
    +{A_{10}}& \: \mathcal{X} \ddot{\mathcal{Y}} +{A_{11}} \: \mathcal{Z} \dot{\mathcal{X}}  +{A_{13}}\: \mathcal{X} \dfrac{\overrightarrow{\nabla}^2 \mathcal{Y}}{a^2} +{A_{14}} \: {\left(\dot{\mathcal{X}}\right)}^{2}+{A_{15}}\: \dfrac{(\overrightarrow{\nabla} \mathcal{X})^2}{a^2} + {A_{16}} \: \dot{\mathcal{X}} \dfrac{\overrightarrow{\nabla}^2 \mathcal{Y}}{a^2} \nonumber \\
    +{A_{17}} & \left. \mathcal{Z} \mathcal{X} +{A_{18}} \: \mathcal{X} \dot{\mathcal{Y}} + {A_{20}} {\mathcal{X}}^{2} \right).
\end{align}

It would be interesting to note that the introduction of new variables in this way~\eqref{gauge_inv_var} does not entail a change in the form of the action~\eqref{three_var_action} when the Newtonian gauge($\beta = E = 0$) is imposed. The choice of variables~\eqref{gauge_inv_var} makes further analysis of the case $A_4 = A_6=0$ (which will be discussed in \ref{A4zero}) noticeably easier. This is due to the fact that the case $A_4 = 0$ is no longer a special one.


\subsection{Integrating out the constraints}\label{A4neq0}

Let us now solve the constraints and analyse the stability of the solution to the equation of motion derived from the action~\eqref{three_var_action}. At this point it is clearly seen \eqref{three_var_action} that the field $\Z$ is non-dynamical and we can derive a $\Z$--constraint which has the following form:
\begin{equation}
\label{Z-constr}
    \Z =  \frac{1}{2 A_3} \left( -A_7 \dfrac{\overrightarrow{\nabla}^2\Y}{a^2}- {A_{8}} \dfrac{\overrightarrow{\nabla}^2\X}{a^2}+ 3 {A_{4}} \dot{\Y} - {A_{11}} \dot{\X} - {A_{17}} \X\right).
\end{equation}
Here we have taken into account that $A_6 = -3 A_4$ (see Appendix A). For now we can notice that the case of $A_3 = 0$ turns constraint~\eqref{Z-constr} to infinity. In what follows, we will take a closer look at this case as well.

So, after substituting the relation~\eqref{Z-constr} into \eqref{three_var_action} and introducing a new variable $\zeta$:
\[\label{zeta}
\zeta = \Y + \eta \X, \quad \eta = \frac{3{A_{11}} {A_{4}}-2{A_{10}} {A_{3}}}{4{A_{1}} {A_{3}}-9{{A_{4}}}^{2}},
\]
we get the already known from unitary gauge analysis result~\cite{Kobayashi:2011nu} (full details can be seen in Appendix C):
\begin{equation}\label{unitary_action}
    S^{(2)} = \int \mathrm{d}t\,\mathrm{d}^3x\,a^3  \left({\mathcal{G}_S \left(\dot{\zeta}\right)}^{2} - \mathcal{F}_S \dfrac{\left(\overrightarrow{\nabla} \zeta\right)^2}{a^2} \right),
\end{equation}
where
\begin{subequations}
    \begin{align}
        \mathcal{G}_S &= \frac{4}{9}\frac{{A_{3}} {{A_{1}}}^{2}}{{A_{4}}^2}-{A_{1}},\\
        \mathcal{F}_S &= -\frac1a \frac{\mathrm{d}}{\mathrm{d}t}\left[\frac{a A_{1} A_{7}}{3 A_{4}}\right]- {A_{2}} = \dfrac{1}{a} \frac{\mathrm{d}}{\mathrm{d}t}\left[ \frac{a A_5 \cdot A_7}{2 A_4}\right] - A_2.
    \end{align}
\end{subequations}
In turn, the square of the sound speed is
\[\label{speed} c_S^2 = c^2_\infty= \frac{\mathcal{F}_S}{\mathcal{G}_S}= 3\frac{{A_{1}} {A_{7}} \dot{{A_{4}}}-3{A_{2}} {{A_{4}}}^{2} -\left({A_{1}} {A_{7}} H+{A_{1}} \dot{{A_{7}}}+{A_{7}} \dot{{A_{1}}}\right) {A_{4}}}{\left(4{A_{1}} {A_{3}}-9{{A_{4}}}^{2}\right) {A_{1}}}.\]

In order to say a few words about the stability analysis we need to use tensor perturbation action~\eqref{tensor_action}. So, to avoid ghost and gradient instabilities we have the same requirements as in unitary gauge analysis:
\[\label{strong_coupling} A_5 > A_2 >0, \quad \mathcal{G}_S > \mathcal{F}_S > 0.\]
{
it follows that 
\[\label{grad} \frac{d \xi}{d t} > a  A_2 > 0,\]
where
\[\label{xi}\xi = \frac{a A_5 \cdot A_7}{2 A_4}.\]
Since $A_4$ is a smooth function of $\pi, H$ and their derivatives it should be finite everywhere. This then implies that $\xi$ can never vanish except at a singularity, $a=0$.}

{Integrating \eqref{grad} from some $t_i$ to $t_f$, we obtain
\[\label{grad2} \xi_f - \xi_i > \int_{t_i}^{t_f} a A_2 \,\mathrm{d} t.\]
Now consider the case of a non-singular universe with $a> \text{const} >0$. The integral in the right-hand side \eqref{grad} can converge if $A_2$ approaches zero very quickly at $+\infty$ or $-\infty$, but this is potentially strong coupling regime in tensor sector, and we do not consider it.}

{Suppose that $\xi_i <0$. Equation \eqref{grad2} reads
\[-\xi_{\mathrm{f}}<\left|\xi_i\right|-\int_{t_i}^{t_{f}} a A_2 \mathrm{~d} t .\]
Since the integral is an increasing function of $t_f$, the righthand side becomes negative for sufficiently large $t_f$. We therefore have $\xi_f>0$, which means that $\xi$ crosses zero.}
In case of Horndeski theory $A_5 = A_7$, so we have $A_5^2$ in the numerator of $\xi$~\eqref{xi}. These two statements contradict each other. However, in the case $A_4=0$ everywhere\cite{Mironov:2019fop} the action~\eqref{unitary_action} coefficients become infinite, which makes it impossible to start the analysis from this point. 

The above leads us to the conclusion that the analysis of the situation when $A_4=0$ everywhere, which also means that $\dot{A_4} = 0$, must be started before all constraints are resolved. This will be the focus of the following sections.

\subsection{$A_4 \equiv 0$ case}\label{A4zero}

It is easy to consider this case if we return to the action~\eqref{three_var_action}, in which we put $A_4 \equiv 0$ (which also mean $\dot{A_4} = 0$). So, after substituting the relation~\eqref{Z-constr} into it and replacing $\Y$ from eq.~\eqref{zeta}, we get the following action:
\begin{align}
\label{A4eq0_action}
    S^{(2)} = \int &\mathrm{d}t\,\mathrm{d}^3x\,a^3 \left( {A_{1}} {\left(\dot{\zeta}\right)}^{2} +{A_{2}} \dfrac{\left(\overrightarrow{\nabla} \zeta\right)^2}{a^2} - \frac{1}{9}\frac{{A_{1}}^2}{{A_{3}}} \dfrac{\left(\overrightarrow{\nabla}^2 \zeta\right)^2}{a^4}+\frac{1}{3} \dfrac{{A_{1}} {A_{11}}}{A_3} \dfrac{\left(\overrightarrow{\nabla}^2 \X \right)}{a^2} \dot{\zeta}\right),
\end{align}
which leads to the constraint $\dot{\zeta} = 0$, which means the absence of dynamics of the field $\zeta$.

This result could have been forseen, since the limit of the sound speed squared~\eqref{speed} equals to zero:
\[\lim_{A_4 \rightarrow 0, \dot{A_4} \rightarrow 0} c^2_S = 0.\]

Thus, in the case where $A_4 = 0$ everywhere, without imposing additional conditions, we obtain a non-dynamical scalar mode. 

\subsubsection{$A_3 = 0$}\label{A3zero}
Let us also distinguish the case $A_3=0$ since from the view of the constraint~\eqref{Z-constr} it can be a singular point. Since the coefficiens are related in the following way 
\begin{align}
    A_3 &= \frac32 A_4 H - \frac12 A_{11} \dot{\pi},
\end{align}
and $A_4 = 0$ already, we have two options: $A_{11} = 0$ and $\dot{\pi} = 0$. Let us consider both cases.

\paragraph{$A_{11} = 0$.}\label{A11zero}
In this case, the $\Z$--constraint gives us the condition:
$$\X =  - \frac{A_{7}}{{A_{8}}} \Y,$$
which brings the action~\eqref{three_var_action} into the following form:
\[S^{(2)} = \int \mathrm{d}t\,\mathrm{d}^3x\,a^3 \,m {\Y}^{2},  \]
where $m = m(A_i)$ is a long combination of  $A_i$ and $\dot{A_i}$ which can be found in Appendix C. This case also turns out to be non-dynamical. 

Now we will proceed to the last case.

\paragraph{$\dot{\pi} = 0$.}\label{dotpizero}

Let us now consider the case of a static background scalar field, i.e. $\dot{\pi} = 0$. In general case, the condition $A_4 = 0$ takes the form of (see Appendix A):
\begin{align} \label{A4}
&K_X\dot{\pi}^3-2G_4H+8HG_{4X}\dot{\pi}^2+8HG_{4XX}\dot{\pi}^4-G_{4\pi}\dot{\pi}-2G_{4\pi X}\dot{\pi}^3& \nonumber\\
&+5H^2G_{5X}\dot{\pi}^3+2H^2G_{5XX}\dot{\pi}^5-3HG_{5\pi}\dot{\pi}^2-2HG_{5\pi X}\dot{\pi}^4&\\\nonumber
&+10HF_4\dot{\pi}^4+4HF_{4X}\dot{\pi}^6+21H^2F_5\dot{\pi}^5+6H^2F_{5X}\dot{\pi}^7=0.
\end{align}
Since $\dot{\pi} = 0$, only the summand $~G_4 H$ remains from the condition~\eqref{A4}.  Because the coefficient $A_2$ is equal to
\[A_2=2G_4-2G_{5X}\dot{\pi}^2\ddot{\pi}-G_{5\pi}\dot{\pi}^2,\]
condition $G_4 = 0$ leads to a strong coupling in the action for tensor perturbations~\eqref{tensor_action}, so it is necessary to impose the condition $H=0$, which corresponds to the case of Minkowski space ($a(t) = \text{const}$).

In this case, the nonzero coefficients of the action~\eqref{three_var_action} remain:
\begin{align}
    {A_{1}} &= -6{G_{4}},~\discretionary{}{}{} {A_{2}} = 2{G_{4}},~\discretionary{}{}{} {A_{8}} = -2G_{4\pi}~\discretionary{}{}{} {A_{13}} = -4G_{4\pi},\nonumber\\
    {A_{14}} &= F_{X}-K_{\pi}, ~\discretionary{}{}{}{A_{15}} = -F_{X}+K_{\pi},~\discretionary{}{}{} {A_{17}} = F_{\pi},~\discretionary{}{}{} {A_{20}} = \frac{1}{2}F_{\pi \pi}.
\end{align}
And the background equations of motion~\eqref{G00back},\eqref{pi_back} impose additional conditions:
$$F = 0, \quad F_{\pi} = 0.$$
Then the action~\eqref{three_var_action} is:
\begin{align}\label{dotpiaction}
    S^{(2)} = \int &\mathrm{d}t\,\mathrm{d}^3x\,a^3 \left(-4{G_{4}} \Z \frac{\overrightarrow{\nabla}^2 \Y}{a^2} - 2G_{4\pi} \Z \frac{\overrightarrow{\nabla}^2 \X}{a^2}-6{G_{4}} {\left(\dot{\Y}\right)}^{2} +\left(F_{X}-K_{\pi}\right) {\left(\dot{\X}\right)}^{2}\right.\nonumber\\
    +& \left.\frac{1}{2}F_{\pi \pi} {\X}^{2}+\left(-F_{X}+K_{\pi}\right) \frac{\left(\overrightarrow{\nabla} \X\right)^2 }{a^2}+2{G_{4}} \frac{\left(\overrightarrow{\nabla} \Y\right)^2 }{a^2}-6G_{4\pi} \dot{\X}\dot{\Y} -4G_{4\pi} \X \frac{\left(\overrightarrow{\nabla} \Y\right)^2 }{a^2}\right).
\end{align}
It follows from the constraint under variation with respect to $\Z$ that
\[\label{Z-constr3}\Y = -\frac12\frac{G_{4 \pi}}{G_{4}} \X.\]
After substituting \eqref{Z-constr3} into \eqref{dotpiaction}, we get:
\begin{align}
    S^{(2)} = \int &\mathrm{d}t\,\mathrm{d}^3x\,a^3 \left( \mathcal{G}_S \left(\dot{\X}\right)^2+ m {\X}^{2} - \mathcal{F}_S \frac{\left(\overrightarrow{\nabla} \X\right)^2 }{a^2} \right),
\end{align}
where
\begin{subequations}
    \begin{align}
        \mathcal{G}_S &= \mathcal{F}_S = \frac{1}{2 G_{4}} \left(3 G_{4 \pi}^2 + 2 F_X G_4 - 2 K_{\pi} G_4\right), \\
        m &= \frac12 F_{\pi \pi}.
    \end{align}
\end{subequations}
And the corresponding equation of motion takes the form:
\[\label{eq_of_motion}\ddot{\X}+ {\frac{G_4 F_{\pi\pi}}{2K_{\pi} G_4-2F_{X} G_4-3 {G_{4\pi}}^{2}}} \X -\frac{\left(\overrightarrow{\nabla} \X\right)^2 }{a^2} = 0\]
This case corresponds to a non-minimally coupled scalar field in Minkowski background. The equation \eqref{eq_of_motion} shows that in this case the speed of sound squared is $c^2_\infty = 1$. However, this does not mean that the scalar mode propagates at the speed of light, since it has mass, but at higher momentum ($k \rightarrow \infty$) the speed of propagation will tend to $c_\infty$. 

\subsection{Brief summary}

Here's a table which is summarizing the results of the previous sections:
\begin{center}
\begin{tabular}{c|c|c}
\hline
    $A_4 \neq 0$ &\multicolumn{2}{c}{$c^2_\infty = {\mathcal{F}_S}/{\mathcal{G}_S}$\eqref{speed}} \\
    \hline
    \multirow{2}{4em}{$A_4\equiv 0$} & $\dot{\pi} \neq 0$ & no dynamics in scalar sector \\
    \cline{2-3}
    &$\dot{\pi} = 0$ & $c^2_\infty = 1$\\ 
    \hline
\end{tabular}
\end{center}

Thus, we obtained that $A_4 = 0$ everywhere, always leads to a stable solution in the scalar perturbation sector. In the case of non-trivial field $\pi$ there are no dynamical scalar perturbations, and thus the stability condition does not arise at all, and in the case of a static background field $\pi$, we obtain a scalar perturbation with the sound speed squared $c_\infty^2 =1$. In the last case, the stability conditions are as follows:
\begin{align}
    &G_4 > 0,&\\
    & 3 G_{4 \pi}^2 + 2 F_X G_4 - 2 K_{\pi} G_4 > 0.&
\end{align}

In the following section we construct specific examples of a bouncing and Genesis scenario in Horndeski theory imposing $A_4 \equiv 0$

\section{Construction of stable solution}\label{solution}

Without loss of generality we choose the following solution for the scalar field
\[
\label{rolling_pi}
\pi (t) = t,
\]
so that $X=1$. Indeed, assuming that the scalar field monotonously increases, one can always obtain~\eqref{rolling_pi} by field redefinition.
To reconstruct the theory which admits arbitrary solution we use the following ansatz for the Lagrangian functions
\begin{subequations}
\label{lagr_func_choice}
\begin{align}
\label{F}
& F(\pi, X) = f_0(\pi) + f_1(\pi)\cdot X, \\
\label{K}
& K(\pi, X) = k_1(\pi) \cdot X, \\
\label{G4}
& G_4(\pi, X) = \frac12,\\
& G_5(\pi, X) = F_4(\pi, X) = F_5(\pi, X) = 0.
\end{align}
\end{subequations}

We are interested to consider the case $G_4 = \text{const}$, which corresponds to GR. Since, as we found out, under the condition $A_4 \equiv 0$ the scalar mode has no dynamics, we do not need to impose stability conditions on the desired solution. Therefore, only the equations of motion and the condition $A_4 \equiv 0$ remain as possible constraints:
\begin{subequations}
\label{lagr_cond}
    \begin{align}
        & {f_{0}}=-\dot{H},& \label{f0} \\
        &{f_{1}} = -3{H}^{2},& \label{f1}\\
        &{k_{1}} = H.& \label{k1}
    \end{align}
\end{subequations}

And then functions~\eqref{lagr_cond} can be directly expressed through the Hubble parameter and its derivative.

Thus, for any Hubble function we can pick the Lagrangian functions describing the behavior of the scalar field in General Relativity. Such a solution will be stable, but the scalar mode will have no dynamics.

\subsubsection*{Bouncing solution}

Hubble parameter can be choosen in the following form for the case of the bounce:
\[\label{H_bounce} H(t)=\frac{t}{3\left(\tau^2+t^2\right)},\]
so that 
\[a(t)=\left(\tau^2+t^2\right)^{\frac{1}{6}},\]
and the bounce occurs at $t= 0$. In what follows we take $\tau \gg 1$ to make this scale safely greater than Planck time. The parameter $\tau$ in \eqref{H_bounce} determines the duration of the bouncing stage.

\begin{figure}[H]\begin{center}\hspace{-1cm}
\label{bounce_pic}
{\includegraphics[width=0.5\linewidth]{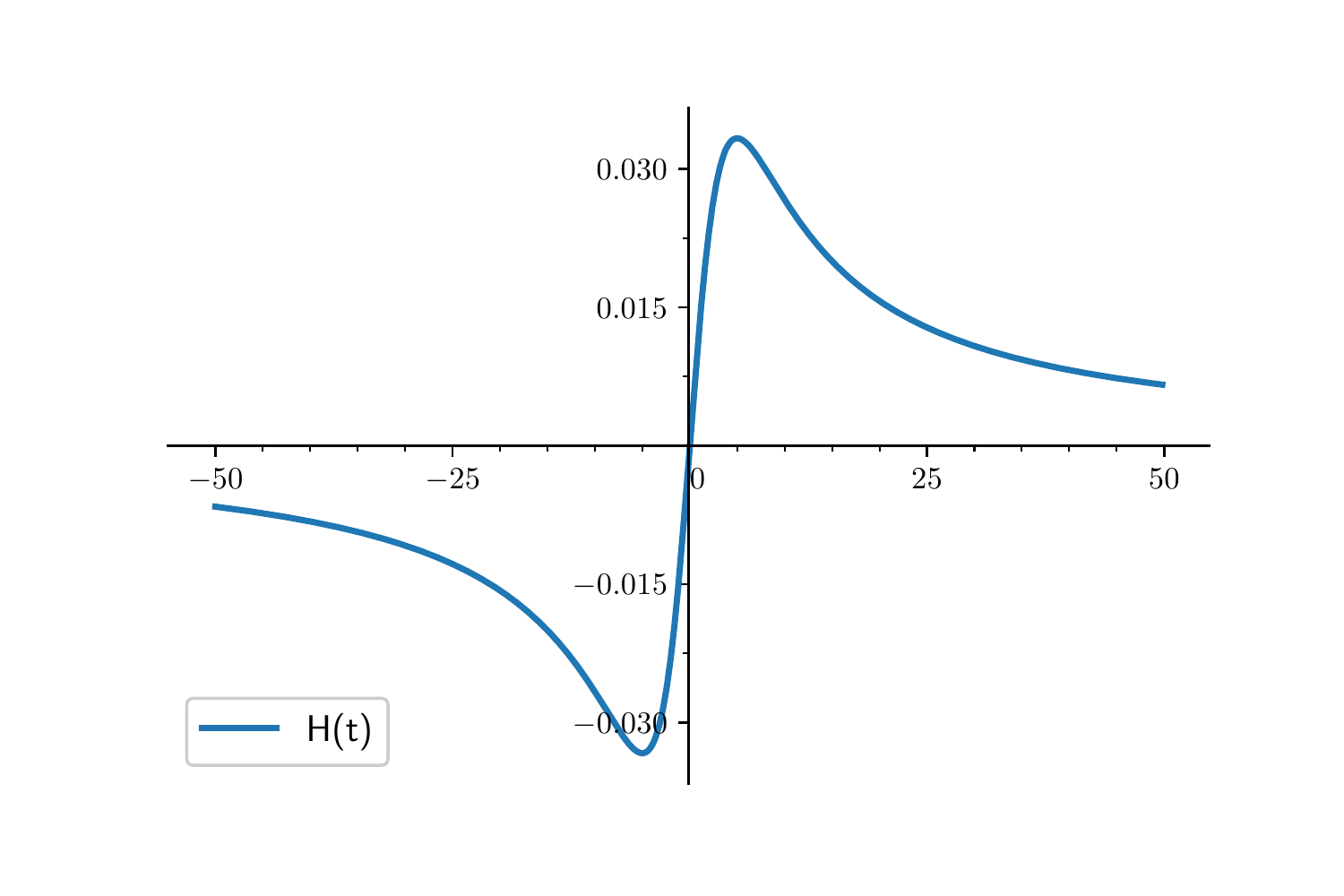}}\hspace{2.8cm}\hspace{-3cm}
{\includegraphics[width=0.5\linewidth]{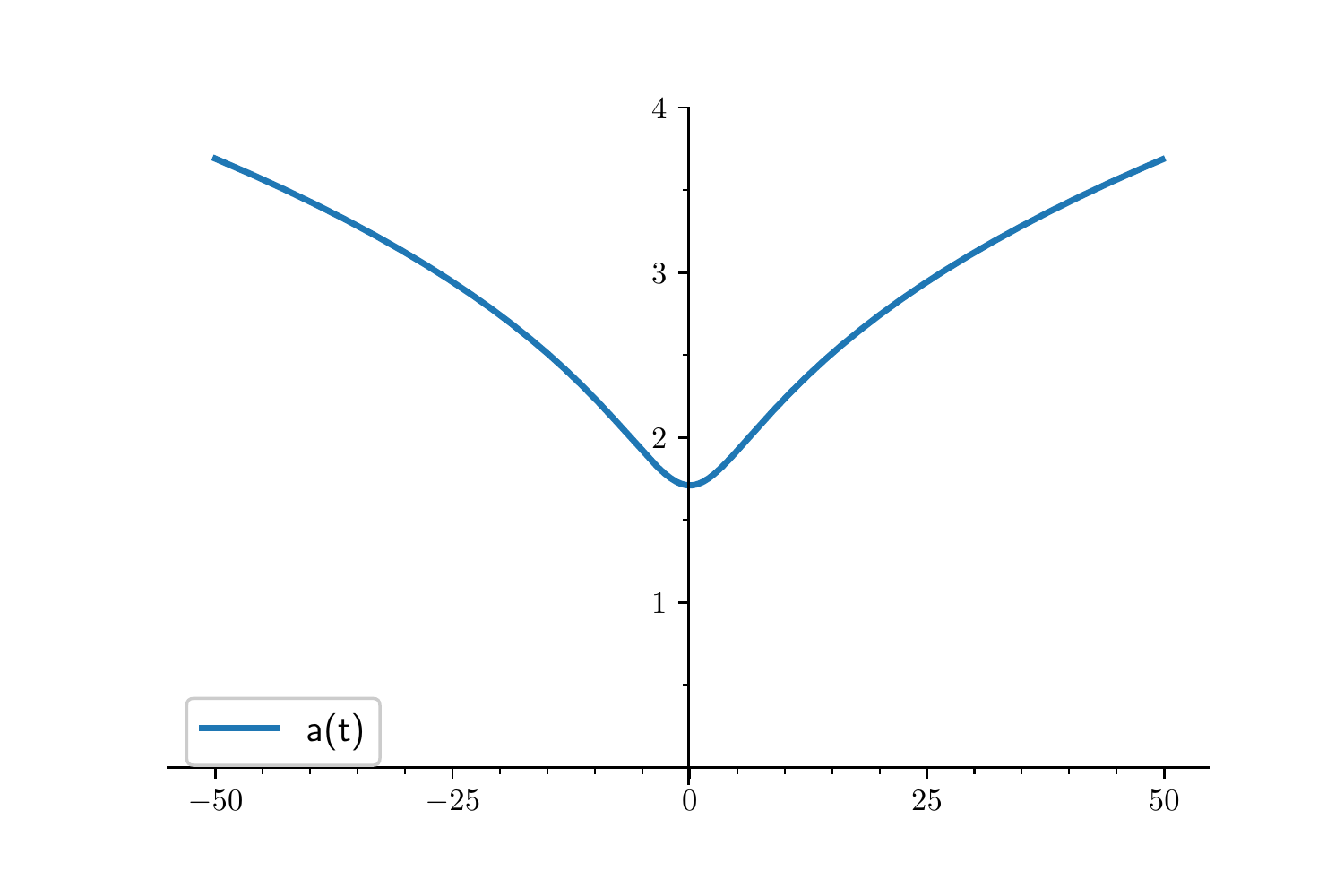}}
{\includegraphics[width=0.5\linewidth]{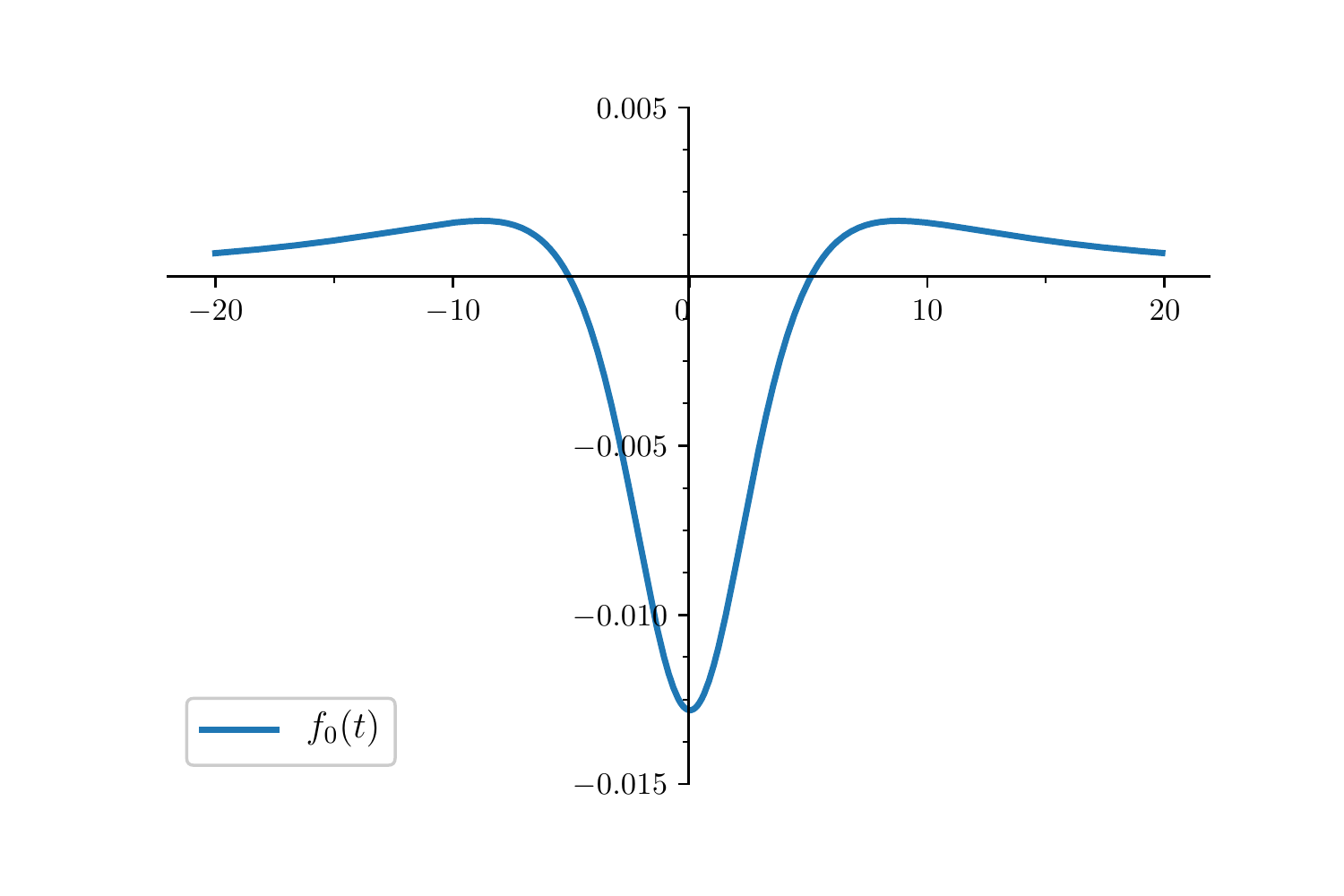}}\hspace{2.8cm}\hspace{-3cm}
{\includegraphics[width=0.5\linewidth]{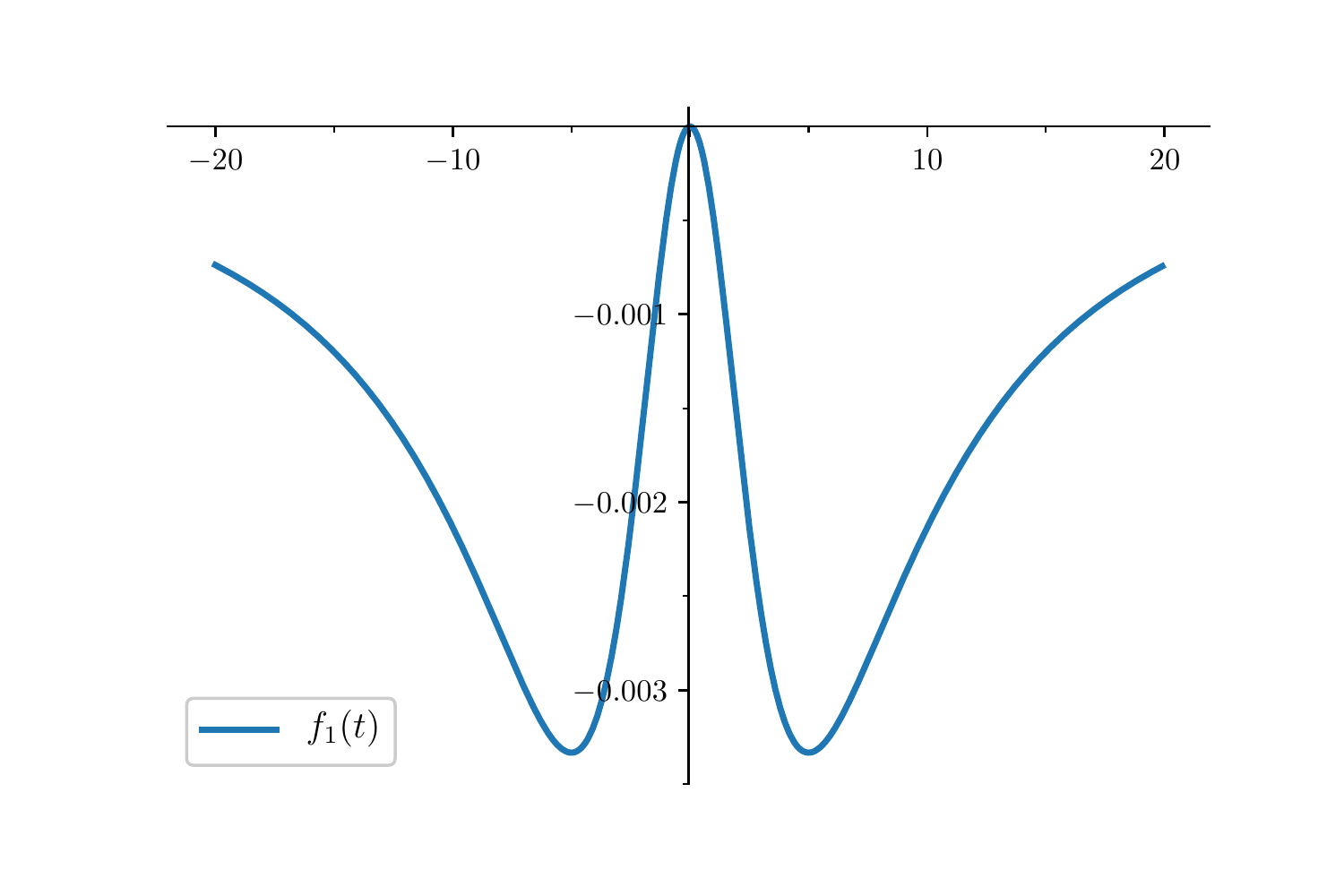}}
\caption{\footnotesize{Hubble parameter $H(t)$, scale factor $a(t)$ and the Lagrangian functions $f_0(t)$, $f_1(t)$ of the bouncing scenario with parameter $\tau = 25$ (recall that $k_1(t) = H(t)$).}} \label{bouncePics}.
\end{center}\end{figure}

We show corresponding scale factor $a(t)$, Hubble parameter and Lagrangian functions in Fig.~\ref{bouncePics}

According to \eqref{lagr_cond} and \eqref{H_bounce} the Lagrangian reads
\[\mathcal{L} = \frac{ \pi^2 - \tau^2}{3\left(\tau^2+\pi^2\right)^2}  - \frac{\pi^2 X}{\left(\tau^2+\pi^2\right)^2}+ \frac{\pi X}{3\left(\tau^2+\pi^2\right)} \Box{\pi}  + \frac12 R.\]
\subsubsection*{Genesis solution}

Genesis case~\cite{Creminelli:2010ba} corresponds to the Hubble parameter with the following asymptotics on $t \rightarrow - \infty$:
\[H(t) \propto \dfrac{1}{(-t)^3}.\]

We consider full evolution which corresponds to a genesis start of the universe with subsequent slowing down to Minkowski space in the end. We choose
\[H(t)=\alpha \frac{\tau^2}{\left(t^2 + \tau^2\right)^{3/2}},\]
where $\alpha$ is an arbitary parameter which is responsible for the ratio of scale factors at $+$ and $-\infty$.
Then the scale factor is
\[a(t) = \exp \left(\frac{\alpha t}{\sqrt{\tau^2 + t^2}} + \alpha\right),\]
\begin{figure}[H]
\begin{center}\hspace{-1cm}
{\includegraphics[width=0.5\linewidth]{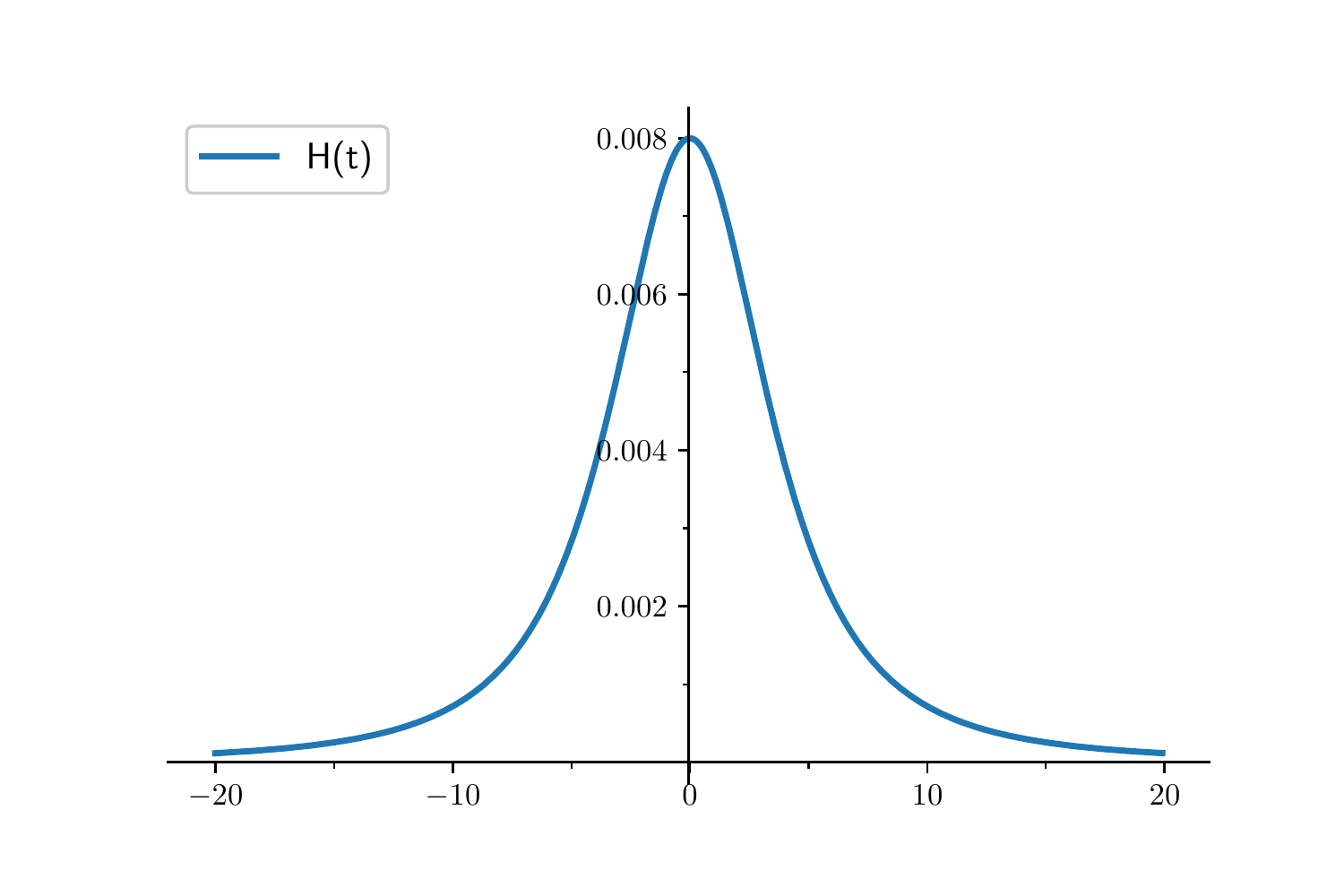}}\hspace{2.8cm}\hspace{-3cm}
{\includegraphics[width=0.5\linewidth]{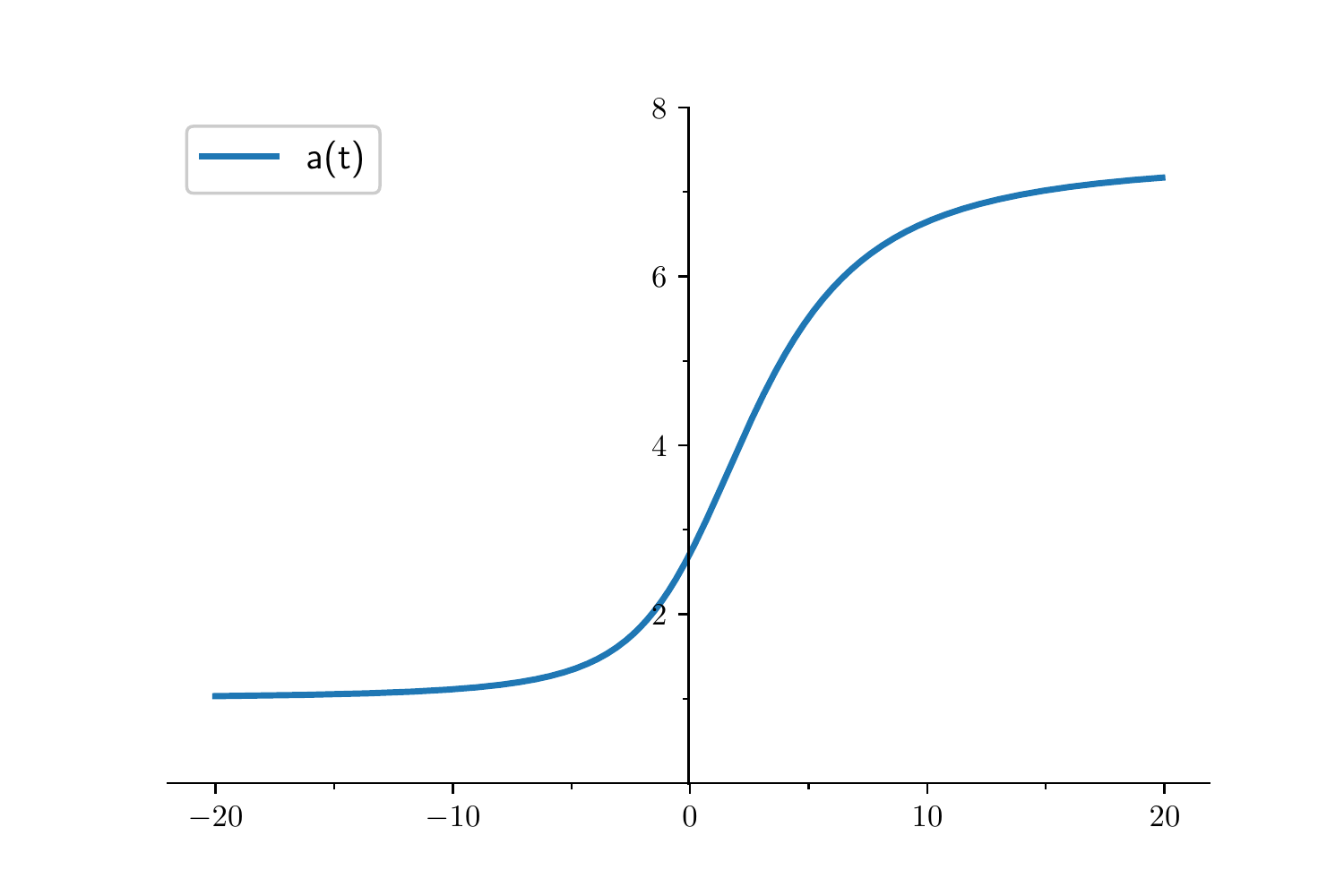}}
{\includegraphics[width=0.5\linewidth]{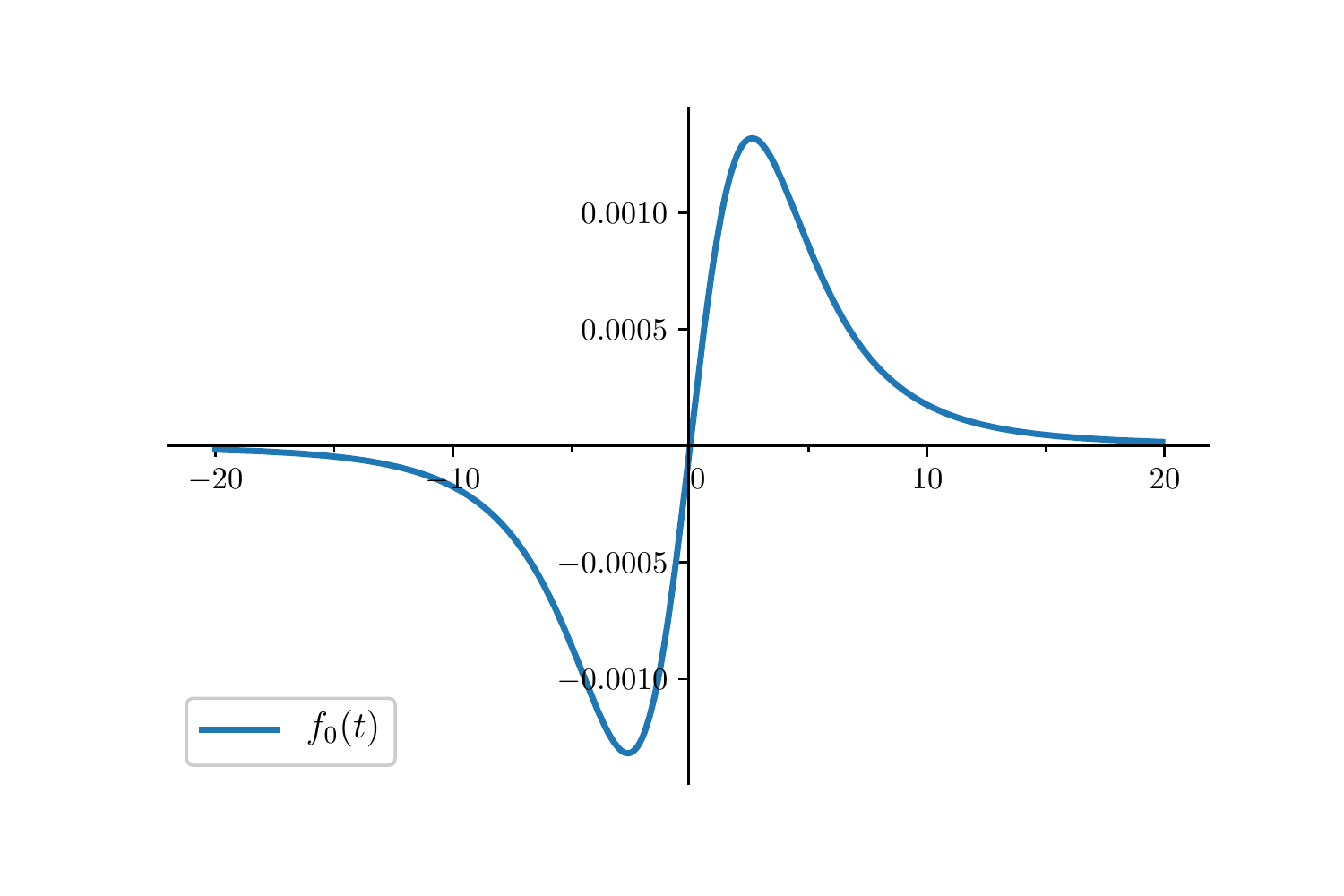}}\hspace{2.8cm}\hspace{-3cm}
{\includegraphics[width=0.5\linewidth]{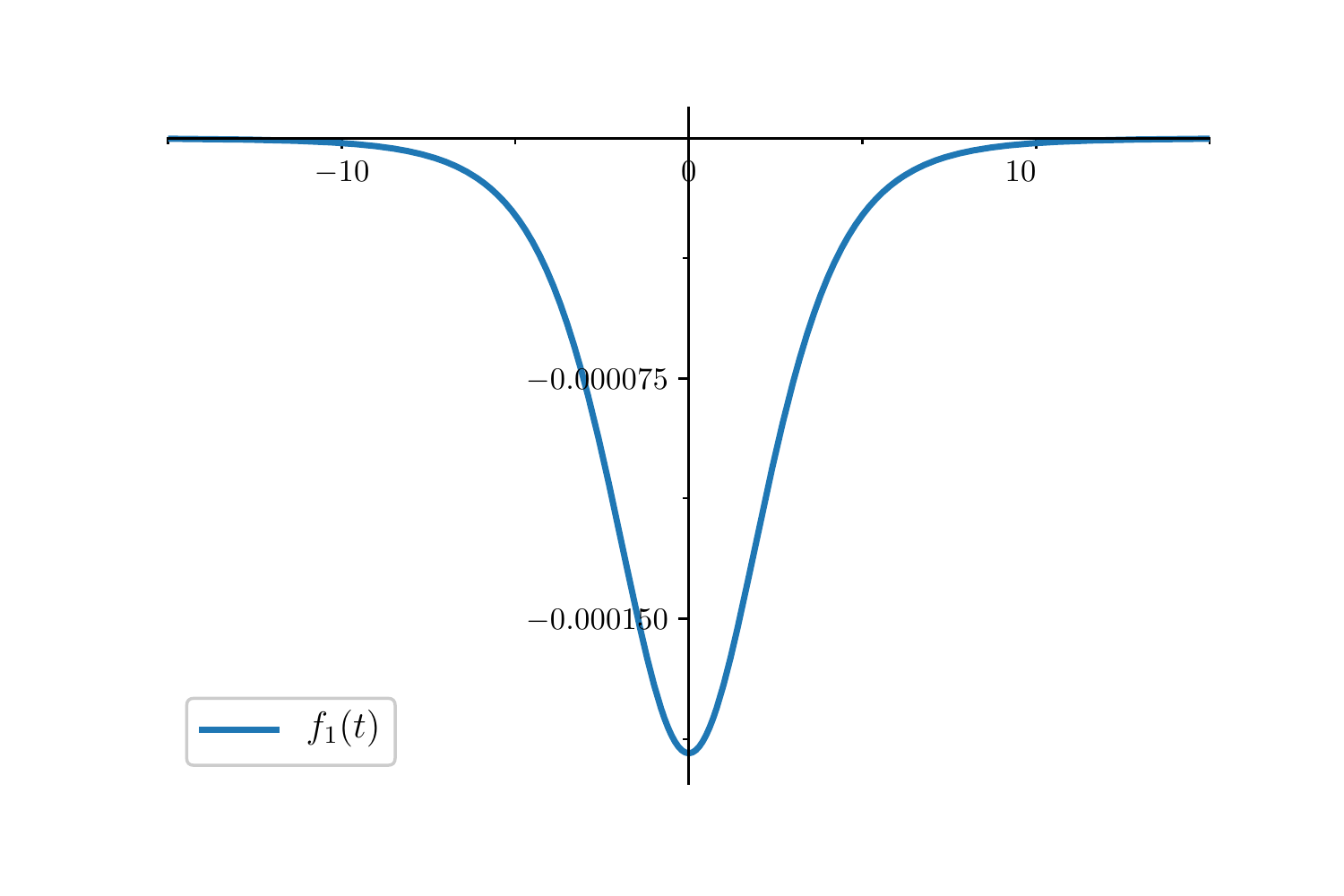}}
\caption{\footnotesize{Hubble parameter $H(t)$, scale factor $a(t)$ and the Lagrangian functions $f_0(t)$, $f_1(t)$ of the Genesis scenario with chose of parameters: $\alpha = 1$, $\tau = 25$ (recall that $k_1(t) = H(t)$).}} \label{GenesisPics}
\end{center}
\end{figure}
which is the solution to the background equations of motion of the Lagrangian:
\[\mathcal{L} = \frac{3 \alpha \tau^2 \pi}{\left(\tau^2+\pi^2\right)^{5/2}} - 3 X \frac{\alpha^2 \tau^4}{\left(\tau^2+\pi^2\right)^3} + X \frac{\alpha \tau^2}{ \left(\tau^2+\pi^2\right)^{3/2}} \Box{\pi} + \frac12 R.\]
The plots of the corresponding Lagrangian functions, the scale factor and the Hubble parameter are shown in Fig.~\ref{GenesisPics}.
\section{Conclusion}

{In this paper we considered a new variant of avoiding of the No-go theorem in Horndeski theory. Making the second-order action explicitly gauge-invariant allowed us to consider the case in which the previously used unitary gauge turned out to be singular. This case is not only interesting from an academic point of view, but also allows us to construct previously unresolved cosmological models, which do not contain singular points throughout evolution. As an example of such models, we considered models of genesis and universe with bounce. The absence of dynamic scalar modes of perturbations can be balanced by the addition of matter. We presume that the source for the scalar perturbations observed in our universe may be additional matter, such as an extra scalar field. However, the main goal of our work was to study the behavior of pure Horndeski theory and find another way to evade No-go theorem.} 

{The case we considered corresponds to the imposition of an additional condition on the model, which is $A_4 = 0$. But does the imposition of such a condition correspond to an unnatural selection of the parameters of the theory?} 

{Since we consider the Lagrangian~\eqref{lagrangian} ($F, K, G_4, G_5$) functions to be arbitrary, the initial restrictions on the metric and the background scalar field are the background equations of motion.} The first Friedman equation~\eqref{G00back} relates $\dot{\pi} = \sqrt{X}$ and $H$ {(i.e., one can get the function $H(\dot{\pi})$)}. Whereas the second Friedman equation~\eqref{Giiback} and {the matter equation~\eqref{pi_back}} 
relate $\ddot{\pi}$, $\dot{H}$, $H$ and $\dot{\pi}$, which allows us to express $\ddot{\pi}$ and $\dot{H}$ as functions of $\dot{\pi}$ and $H$.

Thus, the coefficient $A_4$ is a function of the field $\pi$ and its {first} derivative, which allows us to say that imposing $A_4 \equiv 0$ {does not impose additional restrictions on the initial conditions or solutions of the background equations , but creates a relation for the Lagrangian functions.} But all this reasoning is valid only under the assumption of a homogeneous isotropic background, since otherwise the setup becomes considerably more contrived.

\section*{Acknowledgements}
The authors are grateful to V. Rubakov for useful comments and fruitful discussions and V. Volkova for the careful reading of earlier versions of the text of this paper. {The authors wish to thank Kasper Peeters for developing and maintaining cadabra2 software~\cite{Peeters:2018dyg}, with which most of the calculations were performed.} This work has been supported by Russian Science Foundation grant 19-12-00393.
\newpage
\section*{Appendix A}

In this Appendix we collect the expressions for coefficients $A_i$ 
entering the quadratic actions~\eqref{tensor_action} and \eqref{action}:
\begin{flalign}
\label{A_1}
&A_1=3\left[-2G_4+4G_{4X}\dot{\pi}^2-G_{5\pi}\dot{\pi}^2+2HG_{5X}\dot{\pi}^3 + 2 F_{4}\dot{\pi}^4\ + 6 H F_{5}\dot{\pi}^5\right],&
\end{flalign}
\vspace{-1cm}
\begin{flalign}
\label{A_2}
&A_2=2G_4-2G_{5X}\dot{\pi}^2\ddot{\pi}-G_{5\pi}\dot{\pi}^2,&
\end{flalign}
\vspace{-1cm}
\begin{flalign}\nonumber
\label{A_3}
&A_3=F_X\dot{\pi}^2+2F_{XX}\dot{\pi}^4+12HK_X\dot{\pi}^3+6HK_{XX}\dot{\pi}^5-K_{\pi}\dot{\pi}^2-K_{\pi X}\dot{\pi}^4&\\\nonumber
&-6H^2G_4+42H^2G_{4X}\dot{\pi}^2+96H^2G_{4XX}\dot{\pi}^4+24H^2G_{4XXX}\dot{\pi}^6&\\
&-6HG_{4\pi}\dot{\pi}-30HG_{4\pi X}\dot{\pi}^3-12HG_{4\pi XX}\dot{\pi}^5+30H^3G_{5X}\dot{\pi}^3&\\\nonumber
&+26H^3G_{5XX}\dot{\pi}^5+4H^3G_{5XXX}\dot{\pi}^7-18H^2G_{5\pi}\dot{\pi}^2-27H^2G_{5\pi X}\dot{\pi}^4&\\\nonumber
&-6H^2G_{5\pi XX}\dot{\pi}^6+90H^2F_4\dot{\pi}^4+78H^2F_{4X}\dot{\pi}^6+12H^2F_{4XX}\dot{\pi}^8&\\\nonumber
&+168H^3F_5\dot{\pi}^5+102H^3F_{5X}\dot{\pi}^7+12H^3F_{5XX}\dot{\pi}^9,&
\end{flalign}
\vspace{-1cm}
\begin{flalign}\nonumber
\label{A_4}
&A_4=2\big[K_X\dot{\pi}^3-2G_4H+8HG_{4X}\dot{\pi}^2+8HG_{4XX}\dot{\pi}^4-G_{4\pi}\dot{\pi}-2G_{4\pi X}\dot{\pi}^3&\\
&+5H^2G_{5X}\dot{\pi}^3+2H^2G_{5XX}\dot{\pi}^5-3HG_{5\pi}\dot{\pi}^2-2HG_{5\pi X}\dot{\pi}^4&\\\nonumber
&+10HF_4\dot{\pi}^4+4HF_{4X}\dot{\pi}^6+21H^2F_5\dot{\pi}^5+6H^2F_{5X}\dot{\pi}^7\big],&
\end{flalign}
\vspace{-1cm}
\begin{flalign}
\label{A_5}
&A_5=-\dfrac{2}{3}A_1,&
\end{flalign}
\vspace{-1cm}
\begin{flalign}
\label{A_6}
&A_6=-3A_4,&
\end{flalign}
\vspace{-1cm}
\begin{flalign}
\label{A_7}
&A_7 = -A_5 -A_{16}\dot{\pi},&
\end{flalign}
\vspace{-1cm}
\begin{flalign}\nonumber
\label{A_8}
&A_8=2\big[K_X\dot{\pi}^2-G_{4\pi}-2G_{4\pi X}\dot{\pi}^2+4 HG_{4X}\dot{\pi}+8HG_{4XX}\dot{\pi}^3-2HG_{5\pi}\dot{\pi}&\\
&-2HG_{5\pi X}\dot{\pi}^3+3 H^2G_{5X}\dot{\pi}^2+2 H^2G_{5XX}\dot{\pi}^4 + 10 H F_{4}\dot{\pi}^3
+ 4 H F_{4X}\dot{\pi}^5 &\\\nonumber
&+ 21 H^2 F_{5}\dot{\pi}^4 + 6 H^2 F_{5X}\dot{\pi}^6\big],&
\end{flalign}
\vspace{-1cm}
\begin{flalign}
\label{A_9}
&A_9=- \big(A_8 - A_{16} H\big),&
\end{flalign}
\vspace{-1cm}
\begin{flalign}
\label{A_10}
&A_{10}=-3\big(A_8-A_{16} H\big) ,&
\end{flalign}
\vspace{-1cm}
\begin{flalign}\nonumber
\label{A_11}
&A_{11}=2\big[-F_X\dot{\pi}-2F_{XX}\dot{\pi}^3+K_\pi\dot{\pi}-6HK_{XX}\dot{\pi}^4-9HK_X\dot{\pi}^2+K_{\pi X}\dot{\pi}^3&\\\nonumber
&+3HG_{4\pi}+24HG_{4\pi X}\dot{\pi}^2+12H G_{4\pi XX}\dot{\pi}^4-18H^2G_{4X}\dot{\pi}-72H^2G_{4XX}\dot{\pi}^3&\\
&-24H^2G_{4XXX}\dot{\pi}^5+9H^2G_{5\pi}\dot{\pi}+21H^2G_{5\pi X}\dot{\pi}^3+6H^2G_{5\pi XX}\dot{\pi}^5&\\\nonumber
&-15H^3G_{5X}\dot{\pi}^2-
20H^3G_{5XX}\dot{\pi}^4-4H^3G_{5XXX}\dot{\pi}^6 - 60 H^2 F_{4}\dot{\pi}^3 - 66 H^2 F_{4X}\dot{\pi}^5&\\\nonumber
& - 12 H^2 F_{4XX}\dot{\pi}^7 - 105 H^3 F_{5}\dot{\pi}^4 - 84 H^3 F_{5X}\dot{\pi}^6 - 12 H^3 F_{5XX} \dot{\pi}^8
\big],&
\end{flalign}
\vspace{-1cm}
\begin{flalign}\nonumber
\label{A_12}
&A_{12}=2\big[F_X\dot{\pi}-K_\pi\dot{\pi}+3HK_X\dot{\pi}^2-HG_{4\pi}+G_{4\pi\pi}\dot{\pi}-10HG_{4\pi X}\dot{\pi}^2+6H^2G_{4X}\dot{\pi}&\\
&+12H^2G_{4XX}\dot{\pi}^3-3H^2G_{5\pi}\dot{\pi}+HG_{5\pi\pi}\dot{\pi}^2-4H^2G_{5\pi X}\dot{\pi}^3+3H^3G_{5X}\dot{\pi}^2&\\\nonumber
&+2H^3G_{5XX}\dot{\pi}^4 + 12 H^2 F_{4}\dot{\pi}^3 + 6 H^2 F_{4X}\dot{\pi}^5 - 2 H F_{4\pi}\dot{\pi}^4 + 15 H^3 F_{5} \dot{\pi}^4 + 6 H^3 F_{5X} \dot{\pi}^6 &\\\nonumber
&- 3 H^2 F_{5\pi} \dot{\pi}^5
\big],&
\end{flalign}
\vspace{-1cm}
\begin{flalign}\nonumber
\label{A_13}
&A_{13}= 2\big[4HG_{4X}\dot{\pi} + 4G_{4X}\ddot{\pi} + 8 G_{4XX}\dot{\pi}^2\ddot{\pi} - 2G_{4\pi} +4 G_{4\pi X}\dot{\pi}^2 + 2 H^2G_{5X}\dot{\pi}^2&\\\nonumber
&+2 \dot{H}G_{5X}\dot{\pi}^2+4 H G_{5X}\dot{\pi}\ddot{\pi}+4HG_{5XX}\dot{\pi}^3\ddot{\pi} -2HG_{5\pi}\dot{\pi}-2G_{5\pi}\ddot{\pi}+2HG_{5\pi X}\dot{\pi}^3&\\
&-2G_{5\pi X}\dot{\pi}^2\ddot{\pi}-G_{5\pi\pi}\dot{\pi}^2 + 2 H F_{4}\dot{\pi}^3 + 6 F_{4} \ddot{\pi}\dot{\pi}^2 +4 F_{4X}\ddot{\pi}\dot{\pi}^4 + 2 F_{4\pi}\dot{\pi}^4 + 24 H F_{5} \ddot{\pi}\dot{\pi}^3&\\\nonumber
&+6 H^2 F_{5} \dot{\pi}^4 + 6 \dot{H} F_{5}\dot{\pi}^4 + 12 H F_{5X}\ddot{\pi}\dot{\pi}^5 + 6 H F_{5\pi} \dot{\pi}^5
\big],
\end{flalign}
\vspace{-1cm}
\begin{flalign}\nonumber
\label{A_14}
&A_{14}=F_{X} + 2 F_{XX}\dot{\pi}^2- K_{\pi} + 6 H K_{X}\dot{\pi} -
 K_{\pi X}\dot{\pi}^2 + 6 H K_{XX}\dot{\pi}^3 + 6 H^2 G_{4 X} &\\\nonumber
 &-18 H G_{4\pi X}\dot{\pi} + 48 H^2 G_{4 XX}\dot{\pi}^2 -
 12 H G_{4\pi XX}\dot{\pi}^3 + 24 H^2 G_{4 XXX}\dot{\pi}^4 +
 6 H^3 G_{5 X}\dot{\pi} &\\
 &- 3 H^2 G_{5\pi}-
15 H^2 G_{5\pi X}\dot{\pi}^2 + 14 H^3 G_{5 XX}\dot{\pi}^3 +
 4 H^3 G_{5 XXX}\dot{\pi}^5 - 6 H^2 G_{5\pi XX}\dot{\pi}^4 &\\\nonumber
 &+
 36 H^2 F_{4}\dot{\pi}^2 + 54 H^2 F_{4 X}\dot{\pi}^4 +
 12 H^2 F_{4 XX}\dot{\pi}^6 + 60 H^3 F_{5}\dot{\pi}^3 +
 66 H^3 F_{5 X}\dot{\pi}^5 &\\\nonumber
 &+ 12 H^3 F_{5 XX}\dot{\pi}^7 ,&
\end{flalign}
\vspace{-1cm}
\begin{flalign}\nonumber
\label{A_15}
&A_{15}=-F_{X} - 4 H K_{X}\dot{\pi} - 2 K_{X}\ddot{\pi} + K_{\pi} -
 K_{\pi X}\dot{\pi}^2 - 2  K_{XX}\dot{\pi}^2 \ddot{\pi} -
 6 H^2 G_{4 X}& \\\nonumber
 &- 4 \dot{H} G_{4 X} -
 20 H^2 G_{4 XX}\dot{\pi}^2 - 8 \dot{H} G_{4 XX}\dot{\pi}^2 -
 24 H  G_{4 XX}\dot{\pi}\ddot{\pi} + 12 H  G_{4\pi X}\dot{\pi} \\\nonumber
 &+
 6 G_{4\pi X}\ddot{\pi} - 16 H  G_{4 XXX}\dot{\pi}^3\ddot{\pi} -
 8 H G_{4\pi XX}\dot{\pi}^3 + 4  G_{4\pi XX}\dot{\pi}^2\ddot{\pi} +
 2  G_{4\pi \pi X}\dot{\pi}^2& \\\nonumber
 &- 4 H^3 G_{5 X}\dot{\pi} -
 4 H\dot{H} G_{5 X}\dot{\pi} - 2 H^2  G_{5 X}\ddot{\pi} +
 3 H^2 G_{5\pi} + 2 \dot{H} G_{5\pi} +
 5 H^2  G_{5\pi X}\dot{\pi}^2 &\\
 &+
 2 \dot{H} G_{5\pi X}\dot{\pi}^2 +
 8 H  G_{5\pi X}\dot{\pi}\ddot{\pi} -
 4 H^3  G_{5 XX}\dot{\pi}^3 - 4 H \dot{H} G_{5 XX}\dot{\pi}^3 -
 10 H^2  G_{5 XX}\dot{\pi}^2\ddot{\pi}& \\\nonumber
 &-
 4 H^2  G_{5 XXX}\dot{\pi}^4\ddot{\pi} -
 2 H^2  G_{5\pi XX}\dot{\pi}^4 +
 4 H  G_{5\pi XX}\dot{\pi}^3\ddot{\pi} +
 2 H  G_{5\pi\pi X}\dot{\pi}^3 - 20  F_{4} H^2 \dot{\pi}^2& \\\nonumber
 &-
 10 \dot{H} F_{4}  \dot{\pi}^2 -
 24 H F_{4}\dot{\pi} \ddot{\pi} - 10 H^2  F_{4 X}\dot{\pi}^4 -
 4 \dot{H} F_{4 X}\dot{\pi}^4 -
 36 H  F_{4 X}\dot{\pi}^3\ddot{\pi} - 6 H  F_{4\pi}\dot{\pi}^3& \\\nonumber
 &-
 8 H  F_{4 XX}\dot{\pi}^5\ddot{\pi} -
 4 H  F_{4\pi X}\dot{\pi}^5 - 30 H^3 F_{5} \dot{\pi}^3 -
 36 H\dot{H} F_{5} \dot{\pi}^3 -
 60 H^2 F_{5} \dot{\pi}^2\ddot{\pi} -
 12 H^3 F_{5 X}\dot{\pi}^5& \\\nonumber
 &- 12 H \dot{H} F_{5 X}\dot{\pi}^5 -
 66 H^2  F_{5 X}\dot{\pi}^4\ddot{\pi} -
 12 H^2 F_{5\pi}\dot{\pi}^4 -
 12 H^2  F_{5 XX}\dot{\pi}^6\ddot{\pi} -
 6 H^2  F_{5\pi X}\dot{\pi}^6,&
\end{flalign}
\vspace{-1cm}
\begin{flalign}
\label{A_16}
&A_{16}=4 F_{4} \dot{\pi}^3 + 12 H F_{5}\dot{\pi}^4,&
\end{flalign}
\vspace{-1cm}
\begin{flalign}\nonumber
&A_{17}=F_{\pi}-2 F_{\pi X} \dot{\pi}^2+K_{\pi\pi{}} \dot{\pi}^2-6 H K_{\pi X} \dot{\pi}^3+6 G_{4\pi}  H^2+6 G_{4\pi\pi{}} H \dot{\pi}-24 G_{4\pi X} H^2 \dot{\pi}^2 & \\
&+12 G_{4\pi\pi{}X} H \dot{\pi}^3-24 G_{4\pi XX} H^2 \dot{\pi}^4+9 G_{5\pi\pi{}} H^2 \dot{\pi}^2-10 G_{5\pi X} H^3 \dot{\pi}^3+6 G_{5\pi\pi{}X} H^2 \dot{\pi}^4 &\\\nonumber
&-4 G_{5\pi XX} H^3 \dot{\pi}^5-30 F_{4\pi} H^2 \dot{\pi}^4-12 F_{4\pi X} H^2 \dot{\pi}^6-42 F_{5\pi} H^3 \dot{\pi}^5-12 F_{5\pi X} H^3 \dot{\pi}^7,&
\end{flalign}
\vspace{-1cm}
\begin{flalign}\nonumber
&A_{18}=-6 F_{X} \dot{\pi}+6 K_{\pi} \dot{\pi}-36 H K_{X} \dot{\pi}^2-12 K_{X} \dot{\pi} \ddot{\pi}-6 K_{\pi X} \dot{\pi}^3-12 K_{XX} \dot{\pi}^3 \ddot{\pi}+24 G_{4\pi} H \\\nonumber
&-24 G_{4X} H \ddot{\pi}-108 G_{4X} H^2 \dot{\pi}-24 G_{4X} \dot{H} \dot{\pi}+72 G_{4\pi X} H \dot{\pi}^2-216 G_{4XX} H^2 \dot{\pi}^3\\\nonumber
&-48 G_{4XX} \dot{H} \dot{\pi}^3+36 G_{4\pi X} \dot{\pi} \ddot{\pi}-192 G_{4XX} H \dot{\pi}^2 \ddot{\pi}+24 G_{4\pi XX} \dot{\pi}^3 \ddot{\pi}-96 G_{4XXX} H \dot{\pi}^4 \ddot{\pi}\\\nonumber
&-48 G_{4\pi XX} H \dot{\pi}^4+12 G_{4\pi\pi{}X} \dot{\pi}^3+54 G_{5\pi} H^2 \dot{\pi}+12 G_{5\pi} \dot{H} \dot{\pi}+12 G_{5\pi} H \ddot{\pi}-36 G_{5X} H^2 \dot{\pi} \ddot{\pi}\\\nonumber
&-72 G_{5X} H^3 \dot{\pi}^2-36 G_{5X} H \dot{H} \dot{\pi}^2+6 G_{5\pi\pi{}} H \dot{\pi}^2+42 G_{5\pi X} H^2 \dot{\pi}^3+12 G_{5\pi X} \dot{H} \dot{\pi}^3\\
&+60 G_{5\pi X} H \dot{\pi}^2 \ddot{\pi}-84 G_{5XX} H^2 \dot{\pi}^3 \ddot{\pi}-48 G_{5XX} H^3 \dot{\pi}^4-24 G_{5XX} H \dot{H} \dot{\pi}^4\\\nonumber
&+12 G_{5\pi\pi{}X} H \dot{\pi}^4-12 G_{5\pi XX} H^2 \dot{\pi}^5+24 G_{5\pi XX} H \dot{\pi}^4 \ddot{\pi}-24 G_{5XXX} H^2 \dot{\pi}^5 \ddot{\pi}-216 F_{4} H^2 \dot{\pi}^3\\\nonumber
&-48 F_{4} \dot{H} \dot{\pi}^3-144 F_{4} H \dot{\pi}^2 \ddot{\pi}-36 F_{4\pi} H \dot{\pi}^4-108 F_{4X} H^2 \dot{\pi}^5-24 F_{4X} \dot{H} \dot{\pi}^5-216 F_{4X} H \dot{\pi}^4 \ddot{\pi}\\\nonumber
&-24 F_{4\pi X} H \dot{\pi}^6-48 F_{4XX} H \dot{\pi}^6 \ddot{\pi}-360 F_{5} H^3 \dot{\pi}^4-180 F_{5} H \dot{H} \dot{\pi}^4-360 F_{5} H^2 \dot{\pi}^3 \ddot{\pi}-72 F_{5\pi} H^2 \dot{\pi}^5\\\nonumber
&-144 F_{5X} H^3 \dot{\pi}^6-72 F_{5X} H \dot{H} \dot{\pi}^6-396 F_{5X} H^2 \dot{\pi}^5 \ddot{\pi}-36 F_{5\pi X} H^2 \dot{\pi}^7-72 F_{5XX{}} H^2 \dot{\pi}^7 \ddot{\pi},&
\end{flalign}
\begin{flalign}\nonumber
&A_{19}=3 F_{\pi}-18 F_{X} H \dot{\pi}-6 F_{X} \ddot{\pi}-6 F_{\pi X} \dot{\pi}^2-12 F_{XX} \dot{\pi}^2 \ddot{\pi}+18 H K_{\pi} \dot{\pi}+6 K_{\pi} \ddot{\pi}-54 H^2 K_{X} \dot{\pi}^2\\\nonumber
&-36 H K_{X} \dot{\pi} \ddot{\pi}-18 \dot{H} K_{X} \dot{\pi}^2-36 H K_{XX} \dot{\pi}^3 \ddot{\pi}+3 K_{\pi\pi{}} \dot{\pi}^2-18 H K_{\pi X} \dot{\pi}^3+6 K_{\pi X} \dot{\pi}^2 \ddot{\pi}\\
&+36 G_{4\pi} H^2+18 G_{4\pi} \dot{H}-108 G_{4X} H^3 \dot{\pi}-72 G_{4X} H \dot{H} \dot{\pi}-36 G_{4X} H^2 \ddot{\pi}+108 G_{4\pi X} H^2 \dot{\pi}^2\\\nonumber
&+36 G_{4\pi X} \dot{H} \dot{\pi}^2+108 G_{4\pi X} H \dot{\pi} \ddot{\pi}-288 G_{4XX} H^2 \dot{\pi}^2 \ddot{\pi}-216 G_{4XX} H^3 \dot{\pi}^3-144 G_{4XX} H \dot{H} \dot{\pi}^3\\\nonumber
&-72 G_{4\pi XX} H^2 \dot{\pi}^4+36 G_{4\pi\pi{}X} H \dot{\pi}^3+72 G_{4\pi XX} H \dot{\pi}^3 \ddot{\pi}-144 G_{4XXX} H^2 \dot{\pi}^4 \ddot{\pi}+54 G_{5\pi} H^3 \dot{\pi}\\\nonumber
&+36 G_{5\pi} H \dot{H} \dot{\pi}+18 G_{5\pi} H^2 \ddot{\pi}-36 G_{5X} H^3 \dot{\pi} \ddot{\pi}-54 G_{5X} H^4 \dot{\pi}^2-54 G_{5X} H^2 \dot{H} \dot{\pi}^2+9 G_{5\pi\pi{}} H^2 \dot{\pi}^2\\\nonumber
&+42 G_{5\pi X} H^3 \dot{\pi}^3+36 G_{5\pi X} H \dot{H} \dot{\pi}^3 +90 G_{5\pi X} H^2 \dot{\pi}^2 \ddot{\pi}-84 G_{5XX} H^3 \dot{\pi}^3 \ddot{\pi}-36 G_{5XX} H^4 \dot{\pi}^4\\\nonumber
&-36 G_{5XX} H^2 \dot{H} \dot{\pi}^4 \dot{\pi}^3+18 G_{5\pi\pi{}X} H^2 \dot{\pi}^4-12 G_{5\pi XX} H^3 \dot{\pi}^5+36 G_{5\pi XX} H^2 \dot{\pi}^4 \ddot{\pi}-24 G_{5XXX} H^3 \dot{\pi}^5 \ddot{\pi}\\\nonumber
&-216 F_{4} H^3 \dot{\pi}^3-144 F_{4} H \dot{H} \dot{\pi}^3-216 F_{4} H^2 \dot{\pi}^2 \ddot{\pi}-54 F_{4\pi} H^2 \dot{\pi}^4-108 F_{4X} H^3 \dot{\pi}^5-72 F_{4X} H \dot{H} \dot{\pi}^5\\\nonumber
&-324 F_{4X} H^2 \dot{\pi}^4 \ddot{\pi}-36 F_{4\pi X} H^2 \dot{\pi}^6-72 F_{4XX} H^2 \dot{\pi}^6 \ddot{\pi}-270 F_{5} H^4 \dot{\pi}^4-270 F_{5} H^2 \dot{H} \dot{\pi}^4\\\nonumber
&-360 F_{5} H^3 \dot{\pi}^3 \ddot{\pi}-72 F_{5\pi} H^3 \dot{\pi}^5-108 F_{5X} H^4 \dot{\pi}^6-108 F_{5X} H^2 \dot{H} \dot{\pi}^6-396 F_{5X} H^3 \dot{\pi}^5 \ddot{\pi}\\\nonumber
&-36 F_{5\pi X} H^3 \dot{\pi}^7-72 F_{5 X X{}} H^3 \dot{\pi}^7 \ddot{\pi},&
\end{flalign}
\begin{flalign}\nonumber
&A_{20}=\frac12 F_{\pi\pi{}}-3 F_{\pi X} H \dot{\pi}-F_{\pi X} \ddot{\pi}-F_{\pi\pi{}X} \dot{\pi}^2-2 F_{\pi XX} \dot{\pi}^2 \ddot{\pi}-9 H^2 K_{\pi X} \dot{\pi}^2-3 \dot{H} K_{\pi X} \dot{\pi}^2\\\nonumber
&-6 H K_{\pi X} \dot{\pi} \ddot{\pi}+3 H K_{\pi\pi{}} \dot{\pi}+K_{\pi\pi{}} \ddot{\pi}+\frac12 K_{\pi\pi\pi{}} \dot{\pi}^2-3 H K_{\pi\pi{}X} \dot{\pi}^3+K_{\pi\pi{}X} \dot{\pi}^2 \ddot{\pi}-6 H K_{\pi XX} \dot{\pi}^3 \ddot{\pi}\\\nonumber
&+6 G_{4\pi\pi{}} H^2+3 G_{4\pi\pi{}} \dot{H}-18 G_{4\pi X} H^3 \dot{\pi}-12 G_{4\pi X} H \dot{H} \dot{\pi}-6 G_{4\pi X} H^2 \ddot{\pi}+18 G_{4\pi\pi{}X} H^2 \dot{\pi}^2\\\nonumber
&+6 G_{4\pi\pi{}X} \dot{H} \dot{\pi}^2-36 G_{4\pi XX} H^3 \dot{\pi}^3-48 G_{4\pi XX} H^2 \dot{\pi}^2 \ddot{\pi}-24 G_{4\pi XX} H \dot{H} \dot{\pi}^3-12 G_{4\pi\pi{}XX} H^2 \dot{\pi}^4\\\nonumber
&+18 G_{4\pi\pi{}X} H \dot{\pi} \ddot{\pi}+12 G_{4\pi\pi{}XX} H \dot{\pi}^3 \ddot{\pi}-24 G_{4\pi XXX} H^2 \dot{\pi}^4 \ddot{\pi}+6 G_{4\pi\pi{}\pi X} H \dot{\pi}^3+9 G_{5\pi\pi{}} H^3 \dot{\pi}\\\nonumber
&+6 G_{5\pi\pi{} } H \dot{H} \dot{\pi}+3 G_{5\pi\pi{}} H^2 \ddot{\pi}-6 G_{5\pi X} H^3 \dot{\pi} \ddot{\pi}-9 G_{5\pi X} H^4 \dot{\pi}^2-9 G_{5\pi X} H^2 \dot{H} \dot{\pi}^2+\frac32 G_{5\pi\pi{}\pi} H^2 \dot{\pi}^2\\\nonumber
&+7 G_{5\pi\pi{}X} H^3 \dot{\pi}^3+6 G_{5\pi\pi{}X} H \dot{H} \dot{\pi}^3-6 G_{5\pi XX} H^4 \dot{\pi}^4-6 G_{5\pi XX} H^2 \dot{H} \dot{\pi}^4+15 G_{5\pi\pi{}X} H^2 \dot{\pi}^2 \ddot{\pi}\\\nonumber
&-14 G_{5\pi XX} H^3 \dot{\pi}^3 \ddot{\pi}+6 G_{5\pi\pi{}XX} H^2 \dot{\pi}^4 \ddot{\pi}-2 G_{5\pi\pi{}XX} H^3 \dot{\pi}^5+3 G_{5\pi\pi{}\pi X} H^2 \dot{\pi}^4-4 G_{5\pi XXX} H^3 \dot{\pi}^5 \ddot{\pi}\\\nonumber
&-36 F_{4\pi} H^3 \dot{\pi}^3-24 F_{4\pi} H \dot{H} \dot{\pi}^3-36 F_{4\pi} H^2 \dot{\pi}^2 \ddot{\pi}-9 F_{4\pi\pi{}} H^2 \dot{\pi}^4-18 F_{4\pi X} H^3 \dot{\pi}^5-12 F_{4\pi X} H \dot{H} \dot{\pi}^5\\\nonumber
&-54 F_{4\pi X} H^2 \dot{\pi}^4 \ddot{\pi}-6 F_{4\pi\pi{}X} H^2 \dot{\pi}^6-12 F_{4\pi XX} H^2 \dot{\pi}^6 \ddot{\pi}-45 F_{5\pi} H^4 \dot{\pi}^4-45 F_{5\pi} H^2 \dot{H} \dot{\pi}^4\\
&-60 F_{5\pi} H^3 \dot{\pi}^3 \ddot{\pi}-12 F_{5\pi\pi{}} H^3 \dot{\pi}^5-66 F_{5\pi X} H^3 \dot{\pi}^5 \ddot{\pi}-18 F_{5\pi X} H^4 \dot{\pi}^6-18 F_{5\pi X} H^2 \dot{H} \dot{\pi}^6\\\nonumber
&-6 F_{5\pi\pi{}X} H^3 \dot{\pi}^7-12 F_{5\pi XX}  H^3 \dot{\pi}^7 \ddot{\pi{}}.
\end{flalign}
\begin{flalign}
&A_{21} = - \frac{2}{a^3} \frac{\mathrm{d}}{\mathrm{d} t}\left[A_{12} a^3\right],&\\
&A_{22} = A_8 - A_{16} H,&\\
&A_{23}  = - A_{12}-\frac{1}{a^3}\frac{\mathrm{d}}{\mathrm{d} t}\left[A_9 a^3\right],&\\
&A_{24} = -A_4,\\
&A_{25} = -\frac{1}{a^3}\frac{\mathrm{d}}{\mathrm{d} t}\left[A_4 a^3\right],&
\end{flalign}
\begin{flalign}
&A_{26}  = \frac23 A_1,&\\
&A_{27}  =  \frac23 \frac{1}{a^3}\frac{\mathrm{d}}{\mathrm{d} t}\left[A_1 a^3\right].
\end{flalign}
Note that $A_{16}=0$ in the Horndeski theory.
\section*{Appendix B}
Here we give the background equations of motion:
\begin{subequations}
    \begin{align}
    \label{G00back}
        \delta g^{00}:\quad
    G_{00} = &F-2F_XX-6HK_XX\dot{\pi}+K_{\pi}X+6H^2G_4+6HG_{4\pi}\dot{\pi}
    \\\nonumber&-24H^2X(G_{4X}+G_{4XX}X)+12HG_{4\pi X}X\dot{\pi}
    \\\nonumber&-2H^3X\dot{\pi}(5G_{5X}+2G_{5XX}X)+3H^2X(3G_{5\pi}+2G_{5\pi X}X)
    \\\nonumber&-6H^2X^2(5F_4+2F_{4X}X)-6H^3X^2\dot{\pi}(7F_5+2F_{5X}X)=0,
    \\\nonumber
    \\
    \label{Giiback}
    \delta g^{ij}:\quad
    G_{i i} = &F-X(2K_X\ddot{\pi}+K_\pi)+2(3H^2+2\dot{H})G_4-12H^2G_{4X}X
    \\\nonumber&-8\dot{H}G_{4X}X-8HG_{4X}\ddot{\pi}\dot{\pi}-16HG_{4XX}X\ddot{\pi}\dot{\pi}+2(\ddot{\pi}+2H\dot{\pi})G_{4\pi}
    \\\nonumber&+4XG_{4\pi X}(\ddot{\pi}-2H\dot{\pi})+2XG_{4\pi\pi}-2XG_{5X}(2H^3\dot{\pi}+2H\dot{H}\dot{\pi}+3H^2\ddot{\pi})
    \\\nonumber&-4H^2G_{5XX}X^2\ddot{\pi}+G_{5\pi}(3H^2X+2\dot{H}X+4H\ddot{\pi}\dot{\pi})
    \\\nonumber&+2HG_{5\pi X}X(2\ddot{\pi}\dot{\pi}-HX)+2HG_{5\pi\pi}X\dot{\pi}
    \\\nonumber&-2F_4X(3H^2X+2\dot{H}X+8H\ddot{\pi}\dot{\pi})-8HF_{4X}X^2\ddot{\pi}\dot{\pi}-4HF_{4\pi}X^2\dot{\pi}
    \\\nonumber&-6HF_5X^2(2H^2\dot{\pi}+2\dot{H}\dot{\pi}+5H\ddot{\pi})-12H^2F_{5X}X^3\ddot{\pi}-6H^2F_{5\pi}X^3=0,
\end{align}
\end{subequations}
where $X = \dot{\pi}^2$. The field equation which is obtained by varying over $\pi$ is a linear combination of Friedman equations and their derivatives:
\[\label{pi_back} \delta \pi:\quad \dfrac{4}{\dot{\pi}}\dfrac{d}{dt} \left[G_{00}\right]  - 3 \dfrac{H}{\dot{\pi}} G_{i i} = 0.\] 
\section*{Appendix C}
\subsection*{\textbf{C.1 Four variables action}}
Let us consider the action~\eqref{action} in detail. As mentioned above, it is invariant with respect to the gauge transformations~\eqref{gauge}. Therefore, to begin with, let us introduce a variable that is invariant with respect to spatial gauge transformations:
\begin{equation}
    \kappa = \frac{\beta}{a^2} + \dot{E}.
\end{equation}
In this case, we get:
\begin{align}\label{four_var_action}
    S^{(2)} = \int &\mathrm{d}t\,\mathrm{d}^3x\,a^3 \left({A_{1}}\: {\left(\dot{\Psi}\right)}^{2}+{A_{2}}\: \dfrac{(\overrightarrow{\nabla}\Psi)^2}{a^2}+{A_{3}} \: {\Phi}^{2}+{A_{4}}\: \Phi \dfrac{\overrightarrow{\nabla}^2\kappa}{a^2}+{A_{5}}\: \dot{\Psi} \dfrac{\overrightarrow{\nabla}^2\kappa}{a^2} +{A_{6}}\: \Phi \dot{\Psi} \right.\nonumber \\
    +{A_{7}}& \: \Phi \dfrac{\overrightarrow{\nabla}^2{\Psi}}{a^2}+{A_{8}} \: \Phi \dfrac{\overrightarrow{\nabla}^2{\chi}}{a^2}+{A_{9}}\: \dfrac{\overrightarrow{\nabla}^2{\kappa}}{a^2} \dot{\chi} +{A_{10}} \:\chi\ddot{\Psi}+{A_{11}} \:\Phi \dot{\chi}+{A_{12}}\: \chi \dfrac{\overrightarrow{\nabla}^2\kappa}{a^2}+{A_{13}} \:\chi \dfrac{\overrightarrow{\nabla}^2\Psi}{a^2} \nonumber\\
    +{A_{14}}&\: \left. {\left(\dot{\chi}\right)}^{2}+{A_{15}}\: \dfrac{(\overrightarrow{\nabla}\chi)^2}{a^2} A_{16}\:\dot{\chi}\dfrac{\overrightarrow{\nabla}^2\Psi}{a^2} +{A_{17}}\: \Phi \chi +{A_{18}}\: \chi \dot{\Psi}+{A_{19}}\: \Psi \chi + {A_{20}} \:{\chi}^{2}\right).
\end{align}
Thus, the action~\eqref{four_var_action} is reduced to what is already known, which in other works was obtained by partially fixing the gauge $E = 0$.

\subsection*{\textbf{C.2 Three variables action in general case}}

Let us now introduce three new fully gauge invariant variables:
\begin{align}
    \X &= \chi +  \kappa \dot{\pi} ,\\
    \Y &= \Psi + H \kappa,\\
    \Z &=  \Phi + \dot{\kappa}.
\end{align}
These are the same values as~\eqref{gauge_inv_var}.

After changing the variables the action~\eqref{four_var_action} is reduced to a form that contains only three gauge-invariant variables:
\begin{align}
    S^{(2)} = \int &\mathrm{d}t\,\mathrm{d}^3x\,a^3 \left({A_{1}} \: {\left(\dot{\Y}\right)}^{2} +{A_{2}}\: \dfrac{(\overrightarrow{\nabla}\Y)^2}{a^2} + {A_{3}}\: {\Z}^{2} +{A_{6}}\: \Z \dot{\Y}  +{A_{7}} \: \Z \dfrac{\overrightarrow{\nabla}^2 \Y}{a^2} +{A_{8}} \: \Z \dfrac{\overrightarrow{\nabla}^2 \X}{a^2}\right. \nonumber \\
    +{A_{10}}& \: \X \ddot{\Y} +{A_{11}} \: \Z \dot{\X}  +{A_{13}}\: \X \dfrac{\overrightarrow{\nabla}^2 \Y}{a^2} +{A_{14}} \: {\left(\dot{\X}\right)}^{2}+{A_{15}}\: \dfrac{(\overrightarrow{\nabla} \X)^2}{a^2} + {A_{16}} \: \dot{\X} \dfrac{\overrightarrow{\nabla}^2 \Y}{a^2} \nonumber \\
    +{A_{17}} & \:\left. \Z \X +{A_{18}} \:\X \dot{\Y} + {A_{20}} {\X}^{2} \right),
\end{align}
further we will take into account the interrelation of the coefficients:
\begin{align}
     A_6=-3A_4.
 \end{align}
 The variable $\Z$ is non-dynamical and gives the constraint equation:
\begin{equation}
    \Z =  \frac{1}{2 A_3} \left( -A_7 \dfrac{\overrightarrow{\nabla}^2\Y}{a^2}- {A_{8}} \dfrac{\overrightarrow{\nabla}^2\X}{a^2}+ 3 {A_{4}} \dot{\Y} - {A_{11}} \dot{\X} - {A_{17}} \X\right).
\end{equation}
Let's substitute the $Z$--relation and introduce a new variable:
$$\zeta = \Y\ + \dfrac{D_1}{D_2} \X.$$
This is necessary in order to explicitly identify one dynamic variable and one constraint. As a result, we get an action of the form:
\begin{align}\label{two_var_action}
    S^{(2)} = -\frac14 \int & \mathrm{d}t\,\mathrm{d}^3x\,a^3 \left( \frac1{{A_{3}}} {\left(C \dfrac{(\overrightarrow{\nabla}^2\X)}{a^2}-{A_{7}} \dfrac{(\overrightarrow{\nabla}^2\zeta)}{a^2}\right)}^{2} - \frac{2}{3}\frac{D_3}{A_3}\dfrac{(\overrightarrow{\nabla}^2 \X)}{a^2}\dot{\zeta}\right.\nonumber\\
    -& \left.4\dfrac{(\overrightarrow{\nabla}\zeta)^2}{a^2} \left({A_{2}}+\frac{3}{4 a} \frac{\mathrm{d}}{\mathrm{d}t}\left(\frac{{a A_{4}} {A_{7}}}{{A_{3}}}\right)\right)-\frac{D_2}{A_3}{\left(\dot{\zeta}\right)}^{2}\right),
\end{align}
where
\begin{subequations}
    \begin{align}
        {D_{1}} &= 3{A_{11}} {A_{4}}-2{A_{10}} {A_{3}},\\
        {D_{2}} &= 4{A_{1}} {A_{3}}-9{{A_{4}}}^{2},\\
        {D_{3}} &= 2{A_{1}} {A_{11}}-9{A_{4}} {A_{9}},\\
         \label{C} C &=  {A_{4}} \frac{D_3}{D_2}.
    \end{align}
\end{subequations}
In the process, we took into account that
\[2{A_{16}} {A_{3}}-{A_{11}} {A_{7}}-3{A_{4}} {A_{8}} =  3{A_{4}} {A_{9}} - \frac{2}{3}{A_{1}} {A_{11}}.\]
After varying \eqref{two_var_action} with respect to $\X$ we obtain the constraint equation:
\begin{align}
    \X = \frac{A_7}{C} \zeta - \frac13 \frac{a^2}{k^2} \frac{D_3}{C^2}\dot{\zeta}
\end{align}
Substitute it in the action~\eqref{two_var_action} and we get:
\begin{equation}
    S^{(2)} = \int \mathrm{d}t\,\mathrm{d}^3x\,a^3  \left({\mathcal{G}_S \left(\dot{\zeta}\right)}^{2} - \mathcal{F}_S \dfrac{\left(\overrightarrow{\nabla} \zeta\right)^2}{a^2} \right),
\end{equation}
where
\begin{subequations}
    \begin{align}
        \mathcal{G}_S &= \frac{4}{9}\frac{{A_{3}} {{A_{1}}}^{2}}{{A_{4}}^2}-{A_{1}},\\
        \mathcal{F}_S &= -\frac1a \frac{\mathrm{d}}{\mathrm{d}t}\left[\frac{a A_{1} A_{7}}{3 A_{4}}\right]- {A_{2}} = \dfrac{1}{a} \frac{d}{dt}\left[ \frac{a A_5 \cdot A_7}{2 A_4}\right] - A_2.
    \end{align}
    \end{subequations}
Thus, we obtained the result known from the calculations made in the unitary gauge.
\subsection*{\textbf{C.3} $\boldsymbol{A_4 \equiv 0}$}
The case $A_4=0$ is different because the coefficient $C$~\eqref{C} is zero, which leads to the constraint $\dot{\zeta} = 0$.
\subsubsection*{\textbf{C.3.1} $\boldsymbol{A_{11} = 0}$}

Due to relations:
\begin{align}
    A_3 &= \frac32 A_4 H - \frac12 A_{11} \dot{\pi},\\
    A_{17} & = 3 \dfrac{H}{\dot{\pi}} A_4 - \dfrac{\ddot{\pi}}{\dot{\pi}} A_{11},
\end{align}
action~\eqref{three_var_action} takes the form:
\begin{align}\label{A11zero}
    S^{(2)} &= \int \mathrm{d}t\,\mathrm{d}^3x\,a^3 \left({A_{1}} {\left(\dot{\Y}\right)}^{2}+{A_{2}} \dfrac{\left(\overrightarrow{\nabla} \Y\right)^2}{a^2} + {A_{7}} \Z \dfrac{\left(\overrightarrow{\nabla}^2 \Y\right)}{a^2}+{A_{8}} \Z \dfrac{\left(\overrightarrow{\nabla}^2 \X\right)}{a^2}+{A_{10}} \X\ddot{\Y}\right.\nonumber\\
    &\left.+{A_{13}} \X \dfrac{\left(\overrightarrow{\nabla}^2 \Y\right)}{a^2}+{A_{14}} {\left(\dot{\X}\right)}^{2}+{A_{15}} \dfrac{\left(\overrightarrow{\nabla} \X\right)^2}{a^2}+{A_{16}} \dot{\X} \dfrac{\left(\overrightarrow{\nabla}^2 \Y\right)}{a^2}+{A_{18}} \X \dot{\Y}+{A_{20}} {\X}^{2} \right).
\end{align}
Variation \eqref{A11zero} with respect to $\Z$ leads to the following constraint:
\[\label{A11zero_constr}\X = -\dfrac{A_{7}}{A_8} \Y. \]
We substitute \eqref{A11zero_constr} into \eqref{A11zero} and obtain
\[S^{(2)} = \int \mathrm{d}t\,\mathrm{d}^3x\,a^3 \,m {\Y}^{2},  \]
where
\begin{align}
    m &= \dfrac{1}{a^3}\left(-\dfrac{d^2}{dt^2}\left[\dfrac{{A_{10}} {A_{7}} {a}^{3}}{{A_{8}}}\right]- \dfrac{1}{A_8^4}\dfrac{d}{dt}\left[{a}^{3} {A_{14}} {{A_{7}}}^{2} {A_{8}} \dot{{A_{8}}}\right] +\dfrac{1}{a^3}\dfrac{d}{dt} \left[\dfrac{1}{A_8^3}\right] {A_{14}} {{A_{7}}}^{2} \left(\dot{{A_{8}}}\right) \right.\nonumber\\
    &\left.-\dfrac{d}{dt}\left[{A_{14}} \left(\dot{{A_{7}}}\right) {a}^{3}\right] \dfrac{A_{7}}{{A_{8}^2}}+\frac{1}{2} \dfrac{d}{dt}\left[\dfrac{{A_{7}} {A_{18}} {a}^{3}}{{A_{8}}}\right]+{A_{20}} \dfrac{{A_{7}^2}}{{A^2_{8}}}\right).
\end{align}

\bibliography{horndeski.bib}
\end{document}